\documentclass[twocolumn,twocolappendix]{aastex631}

\usepackage{graphicx}
\usepackage{xcolor,graphicx,xspace,color,longtable}
\usepackage{amssymb,amsmath,shadow,bezier,curves,rotating}
\usepackage{graphicx,chngcntr}
\graphicspath{ {./images/}}
\usepackage{color,soul}
\usepackage{booktabs}
\usepackage[flushleft]{threeparttable}
\usepackage[autostyle=true]{csquotes}
\usepackage[english]{babel}
\usepackage{hyperref}
\usepackage{tabularx}

\newcommand {\h}    {$h^{-1}\,\mathrm{Mpc}$}
\newcommand {\ks}   {$\mathrm{km}~\mathrm{s}^{-1}$}
\newcommand {\hm}   {$h^{-1} \  \mathrm{M}_{\odot}$}

\newcommand {\hmm}   {$h^{-2} \  \mathrm{M}_{\odot}$}

\defcitealias{Biviano16}{Biv16}
\defcitealias{Biviano21}{Biv21}

\DeclareSymbolFont{matha}{OML}{txmi}{m}{it}
\DeclareMathSymbol{\varv}{\mathord}{matha}{118}

\begin{document}

\title[Dynamics of GCLASS and GOGREEN Clusters] {Dynamical Properties and Velocity Dispersion-Mass Relation of $z \sim 1$ Galaxy Clusters from  the GOGREEN and GCLASS Surveys}

\author[0009-0008-3612-8942]{Shrouk Abdulshafy}
\affiliation{Department of Physics, University of California Merced, 5200 North Lake Road, Merced, CA 95343, USA}
\affiliation{Department of Astronomy, Cairo University, 1 Gamaa Street, Giza, 12613, Egypt:~{\href{mailto:shroukabdulshafy@ucmerced.edu}{shroukabdulshafy@ucmerced.edu}}}

\author[0000-0003-3595-7147]{Mohamed H. Abdullah}
\affiliation{Department of Physics, University of California Merced, 5200 North Lake Road, Merced, CA 95343, USA}
\affiliation{Department of Astronomy, National Research Institute of Astronomy and Geophysics, Cairo, 11421, Egypt}

\author[0000-0002-6572-7089]{Gillian Wilson}
\affiliation{Department of Physics, University of California Merced, 5200 North Lake Road, Merced, CA 95343, USA}

\author[0000-0003-4849-9536]{Michael~L.~Balogh}
\affiliation{Department of Physics and Astronomy, University of Waterloo, Waterloo, ON N2L 3G1, Canada}
\affiliation{Waterloo Centre for Astrophysics, University of Waterloo, Waterloo, ON N2L 3G1, Canada}

\author[0009-0003-7202-4159]{Raouf H. Mabrouk}
\affiliation{Department of Astronomy, National Research Institute of Astronomy and Geophysics, Cairo, 11421, Egypt}

\begin{abstract}
We investigate a sample of 14 galaxy clusters from the GOGREEN and GCLASS (GG) spectroscopic datasets within the redshift range \( 0.87 \leq z \leq 1.37 \) and cluster masses $\mathrm{M}_{200} \gtrsim 2\times 10^{14}$ \hm. Using the highly effective GalWeight technique for cluster membership assignment developed by our own team, we derive the dynamical parameters of these clusters through the virial mass estimator. We examine the velocity dispersion-cluster mass relation (\( \sigma \mathrm{MR} \)) for the GG cluster sample. We find, $\log{\sigma_{200}} = (2.94\pm0.02) + (0.37\pm0.09)\log{\mathrm{M}_{200}}$ with an intrinsic scatter of \( \sigma_\mathrm{int} = 0.02 \pm 0.02 \). Our results demonstrate that the \( \sigma \mathrm{MR} \) relation is consistent with predictions from cosmological simulations, highlighting the reliability of the GalWeight technique for cluster membership assignment. Furthermore, the \( \sigma \mathrm{MR} \) validates the robustness of the virial mass estimator in accurately recovering cluster masses and associated parameters. Importantly, our findings confirm that velocity dispersion can be used directly to estimate cluster mass without relying on dynamical mass estimators.
\end{abstract}


\section{Introduction}
Galaxy clusters are the largest gravitationally bound structures in the Universe and are essential in understanding the formation and evolution of cosmic structures \citep{Bahcall88,Postman92,Carlberg96,Sereno12}. They also serve as important tools for testing cosmological models \citep{Haiman01,Reiprich02,Battye03,Abdullah24}. While our understanding has been greatly enhanced by studying nearby clusters, those at high redshifts present a more challenging and yet fascinating frontier in cosmological research. These high-redshift clusters provide valuable insights into early Universe conditions \citep{Overzier16,McConachie22}. Since they are less evolved compared to those at lower redshifts, they offer a unique perspective on the growth and evolution of these massive structures. Their dynamics is influenced by various factors, including the initial conditions of the Universe, the growth of structure through hierarchical merging, and the effects of dark energy and dark matter. Observations of these distant clusters are challenging due to their faintness and the distortions caused by the expanding Universe. However, recent advancements in observational techniques and theoretical models have significantly improved our ability to explore these distant clusters. For instance, simulations of cosmic structure formation, incorporating detailed models of dark matter and gas physics, enhance our understanding of the properties of galaxy clusters at various evolutionary stages \citep{Klypin16,Nelson18}.

The dynamical mass of galaxy clusters is a fundamental quantity in astrophysics and cosmology. It provides a direct measure of the total mass, including dark matter, within a cluster \citep{Voit05, Kravtsov12}. Accurate dynamical mass estimates are crucial for determining the cluster mass function, which is used to constrain cosmological parameters such as \(\Omega_m\) and \(\sigma_8\) \citep{Vikhlinin09,Allen11}. Moreover, they are essential for calibrating mass-observable relations in large photometric surveys, enabling studies of galaxy cluster evolution across cosmic time \citep{Rozo10,Rykoff16}.
Determining the dynamical properties of galaxy clusters, including mass estimates, is highly dependent on accurately identifying the true members of the cluster. The inclusion of interlopers or the exclusion of true members may lead to substantial biases in mass estimates. Various methods have been proposed in the literature to determine cluster membership. Some techniques utilize solely redshift information, while others incorporate both spatial position and redshift data. Methods based on redshift information include the 3$\sigma$-clipping method \citep{Yahil77}, the fixed gapper technique \citep{Beers90, Zabludoff90}, and the jackknife technique \citep{Perea90a}. These methods rely on statistical principles and specific selection criteria to identify cluster members. Methods that incorporate both spatial position and redshift information include the shifting gapper technique \citep{Fadda96} and a technique proposed by \citet{denHartog96} that estimates the escape velocity as a function of distance from the cluster center calculated by the virial mass estimator (e.g., \citealp{Bahcall81, Heisler85}). 

Notably, two advanced techniques are worth mentioning. First, the caustic technique \citep{Diaferio99}, which involves applying the two-dimensional adaptive kernel method (2DAKM, e.g., \citealp{Pisani93, Pisani96}) on galaxies in the phase-space diagram ($r_\mathrm{p}$, $\mathrm{v}_\mathrm{pec}$). Note that $r_\mathrm{p}$ is the projected distance from the cluster center and $\mathrm{v}_\mathrm{pec}$ is the line-of-sight velocity of a galaxy in the cluster frame, calculated as $\mathrm{v}_{\mathrm{pec}} = (\mathrm{v}_\mathrm{obs} - \mathrm{v}_c)/(1 +z_c)$. Here, $\mathrm{v}_\mathrm{obs}$ is the line-of-sight velocity of the galaxy and $z_c$ and $\mathrm{v}_c$ are the cluster redshift and velocity, respectively. The term $(1 + z_c)$ is a correction due to the global Hubble expansion \citep{Danese80} and $c$ is the speed of light. Further details on the implementation of this method can be found in \citet{Serra11}.
Second, the halo-based group finder (\citealp{Yang05, Yang07}), which requires multiple iterations to reliably identify group members, and its application depends on certain assumptions and empirical relations.

The membership and dynamics of the cluster sample under investigation have been previously analyzed by \citet{Biviano16} (hereafter \citetalias{Biviano16}) and \citet{Biviano21} (hereafter \citetalias{Biviano21}). In \citetalias{Biviano16}, they studied 10 clusters at $z \sim 1$. The cluster members were assigned by dividing the redshift distribution of each cluster into groups separated by weighted gaps of $\geq 8$ (see \citealp{Beers90,Beers91}), identifying the group with the highest population as the main cluster. They utilized several algorithms, including those by \citet{denHartog96}, \citet{Fadda96}, \enquote{shifting gapper}, and \citet{Mamon13}, \enquote{Clean}, to establish membership criteria. A galaxy was confirmed as a member if it met the membership criteria of at least three out of the four algorithms employed, including the one used outlined by \citet{Muzzin12} where they required member galaxies to lie within $1500$ km s$^{-1}$ of the mean cluster velocity in the rest frame. 
In their subsequent work, \citetalias{Biviano21} expanded the sample to 14 clusters. They first identified the main cluster peak in redshift space (following \citealp{Beers91,Girardi93}) by selecting galaxies with $c \;| z - z_c | \leq 6000$~\ks, where $z_c$ is the cluster's initial redshift estimate from \citet{Balogh17} for GOGREEN clusters and \citetalias{Biviano16} for GCLASS clusters. Then, they identified main cluster peaks in redshift space and employed the Kernel Mixture Model (KMM) algorithm (\citealp{Mclachlan88,Ashman94}) to analyze the redshift distribution, aiming to detect and separate any merging subclusters. 
Finally, they refined galaxy membership using the Clean and CLUMPS \citepalias{Biviano21} methods, which identify cluster members based on their position in the projected phase-space.
Regarding the dynamics, both \citetalias{Biviano16} and \citetalias{Biviano21} obtained an estimate of the cluster virial radius, $R_{200}$\footnote{The virial radius $R_{200}$ is the radius within which the cluster is in hydrostatic equilibrium. It is approximately equal to the radius at which the density $\rho=\Delta_{200}\rho_c$, where $\rho_c$ is the critical density of the Universe and $\Delta_{200} = 200$ (e.g., \citealp{Carlberg97}).}, from the line-of-sight velocity, $\sigma_\mathrm{los}$, following the iterative \enquote{$\mathtt{MAMPOSSt}$} procedure \citep{Mamon13}. 
This method assumes an NFW model for the mass distribution, the concentration–mass relation of \citet{Gao08}, and the model of \citet{Mamon&Lokas05} for the velocity anisotropy profile, with the same scale radius as that of the mass profile (as suggested by \citealp{Mamon10}). 

Scaling relations are fundamental tools in astrophysics, linking observable properties of astronomical objects to their intrinsic physical characteristics. These empirical relationships, such as those connecting luminosity, temperature, velocity dispersion, or richness to mass, provide an approach to estimate quantities that are otherwise difficult to measure directly. Among these, the velocity dispersion-cluster mass relation ($\sigma$MR), which stands out as particularly effective tool for galaxy clusters. 
This relation not only provides insights into the dynamical state of clusters but also allows for direct estimation of cluster masses from velocity dispersion measurements alone. In many cases, it removes the need for more complex mass modeling or dynamical estimators, making it especially valuable for large surveys and high-redshift studies where detailed information may be limited.

In this paper, we reanalyze the GOGREEN/GCLASS (GG)\footnote{\url{ http://gogreensurvey.ca/data-releases/data-packages/second-public-data-release-dr2/}} cluster sample utilizing GalWeight, a technique for assigning galaxy cluster membership presented in \citet{Abdullah18}. This technique avoids many issues which affect other techniques, e.g., selection criteria, statistical methods, assumption of empirical relations, or the need for multiple iterations. \citet{Abdullah18} demonstrated that GalWeight compares very favorably against four well-known existing cluster membership techniques (shifting gapper, den Hartog, caustic, Spherical Infall Model).
This technique is specifically designed to simultaneously maximize the number of bona fide cluster members while minimizing the number of contaminating interlopers, and can be applied to both massive galaxy clusters and poor galaxy groups. Moreover, it is effective in identifying members in both the virial and infall regions with high efficiency. GalWeight  has been applied to the MDPL2 and Bolshoi N-body simulations and found to be $> 98$\% accurate in correctly assigning cluster membership. 
For the dynamical analysis, we adopt the virial mass estimator (e.g., \citealp{Rines03,Abdullah20a}), and correct for the surface pressure term (e.g., \citealp{The86,Carlberg97}). This procedure was shown to performs very well in comparison to 25 other methods including \enquote{$\mathtt{MAMPOSSt}$}, which were tested and presented in \citet{Old15} and \citet{Abdullah20a}. In this paper, we also present the velocity dispersion–cluster mass relation ($\sigma$MR) for our GG cluster sample. Our results are then compared to those obtained from simulations.

The paper is organized as follows. In Section~\ref{sec:data}, we describe the cluster sample and the Uchuu-UM simulation. In Section~\ref{sec:cls}, we study the membership
as identified using the GalWeight technique and calculate dynamical properties using the virial mass estimator. In Section~\ref{sec:richness}, we study the cluster the scaling 
relation $\sigma$MR. We summarize our conclusions and future work in Section~\ref{sec:conc}. Throughout the paper we adopt $\Lambda$CDM with $\Omega_\mathrm{m}=1-\Omega_\Lambda$, and $H_0=100$ h km s$^{-1}$ Mpc$^{-1}$. Note that throughout the paper we assume $\log{}$ for $\log_{10}$.
\section{Data and Simulation} \label{sec:data}
In this section, we provide an overview of the cluster sample which we utilize in this work. 
The cluster sample analyzed in this study is derived from the GOGREEN (\citealp{Balogh17,Balogh21}) and GCLASS \citep{Muzzin12} surveys\footnote{\url{http://gogreensurvey.ca/data-releases/data-packages/second-public-data-release-dr2/}}. The GOGREEN survey focused on 27 clusters and groups with redshifts between 1.0 and 1.5. Three clusters of this sample were identified by the South Pole Telescope (SPT) survey (\citealp{Brodwin10,Foley11,Stalder13}), and  nine clusters were sourced from the Spitzer Adaptation of the Red-sequence Cluster Survey (SpARCS, \citealp{Muzzin09,Wilson09}). The remaining clusters in the GOGREEN survey were drawn from the COSMOS and Subaru-XMM Deep Survey (\citealp{Finoguenov07,Finoguenov10,George11}). The GCLASS survey covered ten clusters with redshifts between $0.8$ and $1.3$, all from the SpARCS catalog; five of these were also included in the GOGREEN survey. The combined sample from GCLASS and GOGREEN includes $2257$ unique objects with high-quality $z$ measurements (defined by \citealp{Balogh21}): $1529$ from GOGREEN and $728$ from GCLASS. Note that we added 112 unique objects with reliable redshift measurements from the literature \citep{Nantais16, Stalder13, Sifon16, Foley11}, as described in \citet[][Table 1]{Biviano21}. 
We restrict our sample to galaxies with $\mathrm{M}_\ast \geq10^{9.5}~M_\odot$.
This paper examines the same 14 clusters previously analyzed in \citetalias{Biviano21}. Four clusters of this sample are from GOGREEN, five are from GCLASS, and the remaining five are included in both surveys.

The spectroscopic completeness of the GG surveys varies with stellar mass and cluster-centric radius, as discussed in \citet{Balogh21}. Although such completeness  might affect the derived cluster properties, we investigate its impact using simulations and show in \ref{sec:incomp} that no correction is necessary for our analysis.

Both the GG surveys conducted extensive multi-object spectroscopy of galaxies in clusters using the GMOS spectrograph on the Gemini telescopes \citep{Hook04}. The spectroscopic resolving power for these observations was $r = 440$, with an average uncertainty in the radial velocity $cz$ measurements of about 278 \ks. This corresponds to a rest-frame uncertainty of less than 154 \ks~for the clusters. Since this uncertainty is significantly smaller than the typical line-of-sight velocity dispersion ($\sigma_{\text{los}}$) of the GCLASS and GOGREEN clusters, it is sufficiently precise for conducting a robust dynamical analysis of the sample \citepalias{Biviano21}.

\subsection{The Uchuu-UM Galaxy Simulation}

To investigate the spectroscopic completeness of the GG sample, we use data from the Uchuu-UM mock galaxy catalog \citep{Aung23}, derived from the Uchuu cosmological simulation \citep{Ishiyama21}. Uchuu is part of a suite of large, high-resolution $N$-body simulations \citep{Ishiyama21}. The suite includes the largest simulation, Uchuu, which consists of $2.1$ trillion ($12 800^3$) particles in a box of comoving length 2000 \h, with particle mass resolution of $3.27 \times 10^8$ \hm~and gravitational softening length of 4.27 $h^{-1}$ kpc. Uchuu was created using the massively parallel $N$-body TreePM code, \textsc{greem} \citep{Ishiyama09,Ishiyama12}. 
Haloes and subhaloes were identified with \textsc{rockstar} \citep{Behroozi13a} and merger trees constructed with \textsc{consistent trees} \citep{Behroozi13b}. The Uchuu halo mass, $\mathrm{M}_{200}$, is defined as the mass enclosed within an overdensity of $200\rho_c$, where $\rho_c$ is the critical density of the Universe. Halo/subhalo catalogs and their merger trees are publicly available through the Skies \& Universes site.\footnote{\url{http://www.skiesanduniverses.org/Simulations/Uchuu/}}

The Uchuu-UM Galaxy catalog was created using the \textsc{UniverseMachine} model \citep{Behroozi19}, which assigns galaxy properties to dark matter halos based on their assembly histories. The model parametrizes star formation rates as functions of halo mass, growth history, and redshift. Stellar masses are computed by integrating these star formation rates over time, accounting for stellar mass loss. The \textsc{UniverseMachine} parameters were optimized using a Markov Chain Monte Carlo algorithm to match a variety of observational datasets, including stellar mass functions, cosmic star formation rates, and UV luminosity functions, across a wide range of redshifts ($0 < z < 10$). 
In this paper, we analyze the snapshot at redshift $z = 1$ to investigate the completeness of the sample. 
The snapshot at $z = 1$ is selected because it corresponds to the average redshift of the cluster sample.

\section{Dynamical Analysis of Galaxy Clusters}
\label{sec:cls}
The determination of cluster dynamics requires a reliable identification of cluster members. We employ the GalWeight technique \citep{Abdullah18}, which has been validated with both simulations and observations, to define cluster membership. GalWeight identifies the cluster boundary in redshift space with high accuracy, minimizing contamination from interlopers while accurately recovering true cluster members. This provides a robust foundation for subsequent dynamical analyses. For a detailed description of the method and its weighting scheme, we refer the reader to \citet{Abdullah18}. In this section, we describe the procedure that we follow to calculate galaxy cluster masses and their dynamical parameters, such as virial radius, velocity dispersion, and cluster mass.  

In practice, our analysis proceeds in three main steps:  
\begin{enumerate}
    \item Cluster center determination: It is crucial to accurately determine the cluster center, as the distribution of galaxies in the phase–space diagram is highly sensitive to its position. We apply the two-dimensional Adaptive Kernel Method (2DAKM; e.g., \citealp{Pisani96}) to locate the right ascension and declination of the cluster center ($\alpha_c$, $\delta_c$) and the one-dimensional Adaptive Kernel Method (1DAKM) to determine the redshift ($z_c$). This ensures a robust definition of the cluster center and allows us to construct the line-of-sight velocity ($v_{\mathrm{pec}}$) versus projected radius ($R_p$) phase–space diagram.  
    \item Membership assignment: The GalWeight technique is applied to separate cluster galaxies from foreground and background interlopers in projected phase-space.  
    \item Dynamical analysis: Using the identified members, we compute the velocity dispersion, virial mass, and other dynamical quantities of each cluster. 
\end{enumerate}

The cluster mass can be estimated from the virial mass estimator (e.g., \citealp{Limber60,Binney87,Rines03}) and NFW mass profile \citep{NFW96,NFW97} as follows. 
{The virial mass estimator is given by
\begin{equation} \label{eq:vir1}
M_{VT}=\frac{3\pi N \sum_{i}v_{\mathrm{pec}, i}^2}{2G\sum_{i\neq j}\frac{1}{R_{ij}}} 
\end{equation}
\noindent where $v_{\mathrm{pec,i}}$ is the galaxy line-of-sight velocity in the cluster frame, calculated as $v_{\mathrm{pec}} = ({v}_\mathrm{obs} - v_c)/(1 +z_c)$
and $R_{ij}$ is the projected distance between two galaxies. Here, $v_\mathrm{obs}$ is the line-of-sight velocity of the galaxy and $z_c$ and $v_c$ are the cluster redshift and velocity, respectively. Equation~\ref{eq:vir1} holds only inside the virialized region, which extends to about $R_{200}$.

Since galaxy clusters extend beyond the virial radius, Equation~(\ref{eq:vir1}) tends to overestimate the mass due to the contribution of external pressure from material outside the virialized region \citep{The86,Carlberg97,Girardi98a}. The corrected virial mass is therefore given by:
\begin{equation} \label{eq:vir2}
M_{VTC}=M_{VT}[1-S(r)],
\end{equation}
\noindent where $S(r)$ is a term introduced to correct for surface pressure. 

For an NFW density profile with isotropic orbits (i.e., where the projected velocity dispersion $\sigma_v$ and the tangential component $\sigma_\theta$ are equal, or equivalently where the anisotropy parameter $\beta = 1 - \sigma_\theta^2/\sigma_r^2 = 0$), $S(r)$ can be calculated by
\begin{equation}\label{eq:vir_25}
S(r)=\left(\frac{x}{1+x}\right)^2\left[\ln(1+x)-\frac{x}{1+x}\right]^{-1}\left[\frac{\sigma_v(r)}{\sigma(<r)}\right]^2,
\end{equation}
\noindent where $x=r/r_s$, $r_s$ is the scale radius, $\sigma(<r)$ is the integrated three-dimensional velocity dispersion within $r$, and $\sigma_v(r)$ is a projected velocity dispersion (e.g., \citealp{Koranyi00,Abdullah11}).

The mass density within a sphere of radius $r$ introduced by NFW is given by
  \begin{equation} \label{eq:NFW1}
  \rho(r)=\frac{\rho_s}{x\left(1+x\right)^2}, \end{equation}
\noindent and its corresponding mass is given by
   \begin{equation} \label{eq:NFW11}
   M(<r)=\frac{M_s}{\ln(2)-(1/2)}\left[\ln(1+x)-\frac{x}{1+x}\right],
   \end{equation}
\noindent where $M_s=4\pi\rho_s r^3_s [\ln(2)-(1/2)]$ is the mass within $r_s$, $\rho_s = \delta_s \rho_c$ is the characteristic density within $r_s$ and $\delta_s = (\Delta_{200}/3) c^3 \left[\ln(1+c) - \frac{c}{1+c}\right]^{-1}$, and the concentration $c = R_{200}/r_s$  (e.g.,  \citealp{NFW97,Rines03,Mamon13}). 

The projected number of galaxies within a cylinder of radius R is given by integrating the NFW profile (Equation~\ref{eq:NFW1}) along the line of sight (e.g., \citealp{Bartelmann96,Zenteno16})
   \begin{equation} \label{eq:NFW2}
   N(<R)=\frac{N_s}{\ln(2)-(1/2)} g(x),
   \end{equation}
\noindent where $N_s$ is the number of galaxies within $r_s$ that has the same formula as $\mathrm{M}_s$, and $g(x)$ is given by (e.g., \citealp{Golse02,Mamon&Boue10})
   \begin{equation}
   g(x) = \begin{cases} \ln(x/2) + \frac{\cosh^{-1} (1/x)}{\sqrt{1-x^2}} \ \ \ \mbox{if} \  x \ < \ 1\\ 
	                      1-\ln(2)  \ \ \ \ \ \ \ \ \ \ \ \ \ \ \ \ \ \  \  \mbox{if} \  x \ = \ 1 \\ 
	                      \ln(x/2) + \frac{\cos^{-1} (1/x)}{\sqrt{x^2-1}}  \ \  \ \ \ \mbox{if} \  x \ > \ 1
   \end{cases} \end{equation}

Note that the virial theorem is valid within the virialized region. In our approach,  we first fit the NFW scale radius $r_s$ to calculate $S(r)$. Then, we use Equations~\ref{eq:vir1}–\ref{eq:vir2} to determine $R_{200}$ by requiring that the mean enclosed density equals $200\,\rho_c$. The cluster virial mass is then defined as the corrected virial mass $M_{VTC}$ enclosed within this radius.

\begin{figure*}\hspace{-0cm}    \includegraphics[width=0.75\linewidth]{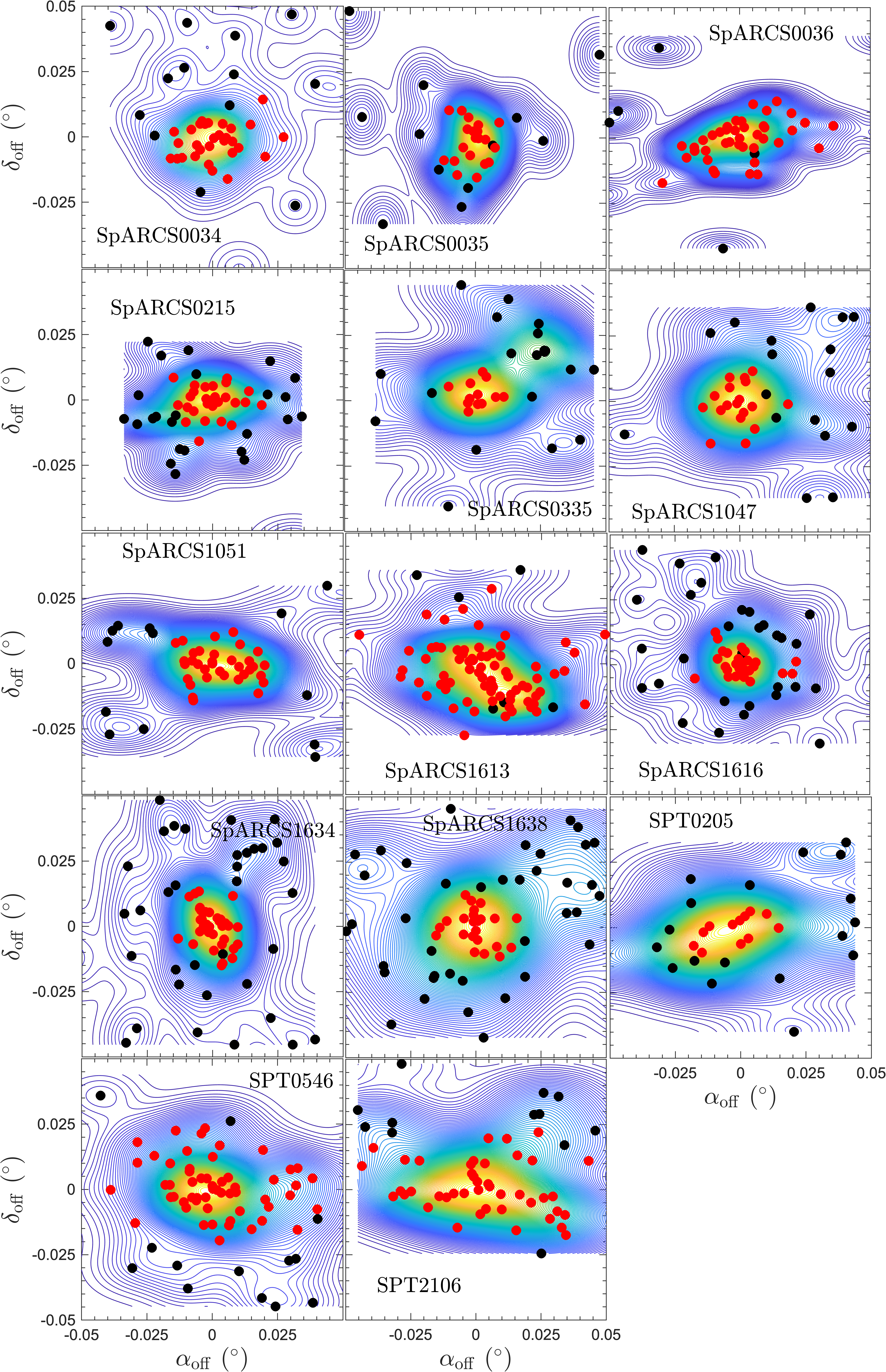} \vspace{-.0cm}
\centering
    \caption{The black points show offsets ($\alpha_\mathrm{off}$, $\delta_\mathrm{off}$) from the cluster center for all spectroscopically-confirmed galaxies (members and non-members) within a velocity gap of $\Delta v \pm 3500$~km~s$^{-1}$ of each cluster redshift. The red points show spectroscopically-confirmed members which satisfy these criteria and also lie within $R_{200}$ (see also Fig~\ref{fig:PS} which shows the location of red and black galaxies in projected phase space). The contours were calculated using the two-dimensional adaptive kernel method (2DAKM), and show the density of (red and black) spectroscopically-confirmed galaxies.}
   \label{fig:Coors}
\end{figure*}

\begin{figure*}\hspace{-0cm}  \includegraphics[width=0.75\linewidth]{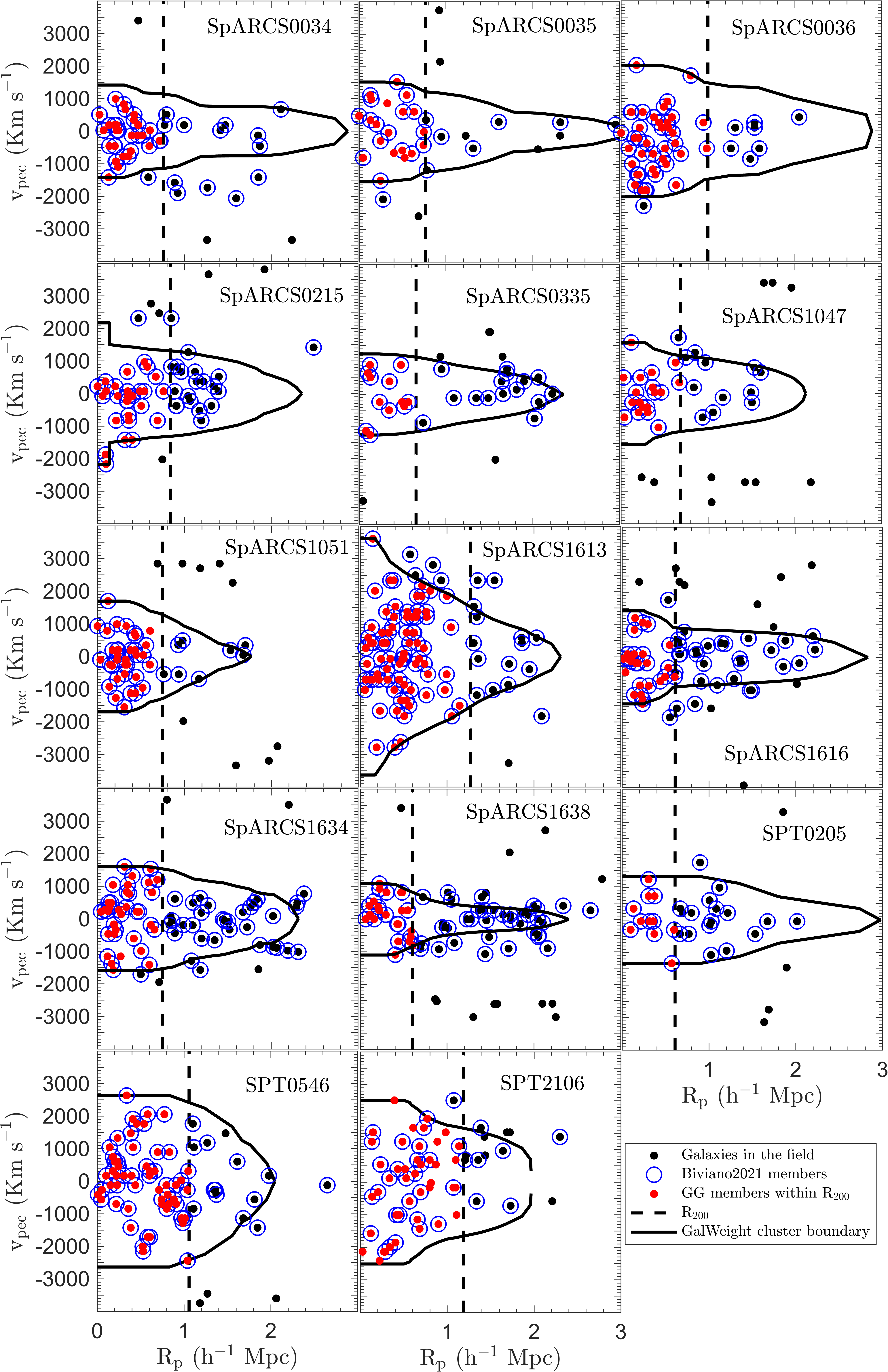} \vspace{-0.1cm}
\centering
     \caption{Projected phase-space diagrams where $\mathrm{v_{pec}}$ and $r_p$ are the peculiar velocity and the projected distance, respectively. Application of the GalWeight technique results in the solid black line (cluster boundary) within which are enclosed spectroscopically-confirmed galaxies which are identified as cluster members. Black points show all spectroscopically-confirmed galaxies within the field of view (shown also in Fig~\ref{fig:Coors}) and red points indicate cluster members identified by the GalWeight technique (black contour) but which also fall within $R_{200}$ (dashed vertical line). Blue circles denote cluster members identified by \citetalias{Biviano21}.}
   \label{fig:PS}
\end{figure*}

\begin{table} \centering
\caption{Coordinates and Redshifts of the GG clusters.}
\label{tab:pos}
\begin{tabular}{lrrrcc}
\hline
&\multicolumn{3}{c}{Results from this work}&\multicolumn{1}{c}{\citetalias{Biviano16}}&\multicolumn{1}{c}{\citetalias{Biviano21}}\\
\cmidrule(lr){2-4}\cmidrule(lr){5-5}\cmidrule(lr){6-6}
\multicolumn{1}{c}{Cluster ID}   &\multicolumn{1}{c}{$\alpha_c$}  &\multicolumn{1}{c}{$\delta_c$} &\multicolumn{1}{c}{$z_c$ }  &\multicolumn{1}{c}{$z_c$}&\multicolumn{1}{c}{$z_c$}\\
&\multicolumn{1}{c}{(deg)}     &\multicolumn{1}{c}{(deg)}&&& \\
\hline

\multicolumn{6}{c}{SpARCS clusters}\\
\hline
SpARCS0034	       &8.6728		&-43.1322	 &0.867   &0.866    &0.867\\
SpARCS0035	       &8.9556	    &-43.2065	 &1.335   &1.336    &1.336\\
SpARCS0036	       &9.1877		&-44.1805	 &0.871   &0.869    &0.870\\
SpARCS0215	       &33.8491	    &-3.7266	 &1.004   &1.004    &1.004\\
SpARCS0335         &53.7669     &-29.4829    &1.369   &  - - -       &1.369\\
SpARCS1047	       &161.8877    &57.6913	 &0.956   &0.956    &0.957\\
SpARCS1051	       &162.7812	&58.3020	 &1.035   &1.034    &1.034\\
SpARCS1613	       &243.3100	&56.8283	 &0.870   &0.872    &0.870\\
SpARCS1616	       &244.1723	&55.7549	 &1.157   &1.155    &1.156\\
SpARCS1634         &248.6516	&40.3618	 &1.177   &1.177    &1.178\\
SpARCS1638         &249.7143	&40.6429	 &1.194   &1.195    &1.194\\
\hline
\multicolumn{6}{c}{SPT clusters}\\
\hline
SPT0205            &31.4471     &-58.4864    &1.322   &  - - -     &1.323\\
SPT0546            &86.6551     &-53.7584    &1.067   &  - - -     &1.067\\
SPT2106            &316.5084    &-58.7482    &1.127   &  - - -     &1.131\\

\hline
\end{tabular}
\end{table} 

\begin{table*} \centering
\caption{Dynamical parameters of the GG cluster sample.}
\label{tab:dyn}
\begin{tabular}{lccrrcrc} \hline
&\multicolumn{4}{c}{Results from this work}&\multicolumn{3}{c}{\citetalias{Biviano21}}\\          
\cmidrule(lr){2-5}\cmidrule(lr){6-8}  \multicolumn{1}{l}{ID}      &\multicolumn{1}{c}{$R_{200}$}       &\multicolumn{1}{c}{$\mathrm{N_{200}}$} &\multicolumn{1}{c}{$\sigma_{200}$}  &\multicolumn{1}{c}{$\mathrm{M}_{200}$}  &\multicolumn{1}{c}{$R_{200}$}     &\multicolumn{1}{c}{$\sigma_{los}$}            &\multicolumn{1}{c}{$\mathrm{M}_{200}$}\\
&\multicolumn{1}{c}{($h^{-1}$~Mpc)}   &           &\multicolumn{1}{c}{(km s$^{-1}$)}                &\multicolumn{1}{c}{($10^{14}~h^{-1}~\mathrm{M}_{\odot}$)} &\multicolumn{1}{c}{($h^{-1}$~Mpc)}    &\multicolumn{1}{c}{(km s$^{-1}$)}        &\multicolumn{1}{c}{($10^{14}~h^{-1}~\mathrm{M}_{\odot}$)}\\     
\hline
\multicolumn{8}{c}{SpARCS clusters}\\
\hline 
SpARCS0034 & 0.681 & 27 & $611_{-116}^{+155}$ & $1.98 \pm 0.31$ & 0.406 & $405\pm51$ & 0.42$\pm$0.14 \\
SpARCS0035 & 0.759 & 19 & $813_{-160}^{+247}$ & $4.70 \pm 0.83$ & 0.630 & $840\pm111$ & 2.66$\pm$1.05 \\
SpARCS0036 & 1.009 & 38 & $934_{-167}^{+245}$ & $6.48 \pm 0.90$ & 0.742 & $799\pm82$ & 2.52$\pm$0.77 \\
SpARCS0215 & 0.823 & 27 & $780_{-190}^{+256}$ & $4.11 \pm 0.65$ & 0.616 & $656\pm70$ & 1.68$\pm$0.56 \\
SpARCS0335 & 0.678 & 12 & $735_{-183}^{+233}$ & $3.48 \pm 0.71$ & 0.483 & $542\pm75$ & 1.26$\pm$0.49 \\
SpARCS1047 & 0.685 & 18 & $651_{-169}^{+276}$ & $2.24 \pm 0.40$ & 0.637 & $668\pm89$ & 1.75$\pm$0.70 \\
SpARCS1051 & 0.732 & 32 & $690_{-122}^{+154}$ & $3.00 \pm 0.45$ & 0.588 & $689\pm75$ & 1.54$\pm$0.49 \\
SpARCS1613 & 1.271 & 76 & $1300_{-161}^{+232}$ & $12.94 \pm 1.39$ & 1.078 & $1185\pm90$ & 7.77$\pm$1.75 \\
SpARCS1616 & 0.669 & 31 & $683_{-129}^{+168}$ & $2.64 \pm 0.40$ & 0.644 & $782\pm71$ & 2.31$\pm$0.63 \\
SpARCS1634 & 0.755 & 31 & $892_{-155}^{+182}$ & $3.88 \pm 0.58$ & 0.595 & $715\pm60$ & 1.89$\pm$0.49 \\
SpARCS1638 & 0.588 & 19 & $541_{-148}^{+267}$ & $1.86 \pm 0.33$ & 0.511 & $564\pm53$ & 1.19$\pm$0.35 \\
\hline
\multicolumn{8}{c}{SPT clusters}\\
\hline
SPT0205  & 0.597 & 10 & $714_{-265}^{+428}$ & $2.259 \pm 0.488$ & 0.595 & $678\pm91$ & 2.17$\pm$0.84 \\
SPT0546  & 1.058 & 55 & $1061_{-166}^{+211}$ & $9.408 \pm 1.143$ & 0.805 & $977\pm84$ & 4.06$\pm$1.05 \\
SPT2106  & 1.161 & 41 & $1240_{-196}^{+238}$ & $13.326 \pm 1.805$ & 0.847 & $1055\pm106$ & 5.11$\pm$1.54 \\
\hline
\end{tabular}
\end{table*}
\subsection{Results} \label{sec:results}
 The black points in Figure~\ref{fig:Coors} show offsets from the cluster center ($\alpha_\mathrm{off}$, $\delta_\mathrm{off}$) for all spectroscopically-confirmed galaxies (members and non-members) within a velocity gap of $\Delta v \pm 3500$~km~s$^{-1}$ of each cluster redshift. The red points show spectroscopically-confirmed members which satisfy these criteria and also lie within $R_{200}$ (see also Figure~\ref{fig:PS} which shows the location of red and black galaxies in projected phase space). The contours were calculated using the two-dimensional adaptive kernel method (2DAKM), and show the density of (red and black) spectroscopically-confirmed galaxies. 
Table~\ref{tab:pos} lists the ($\alpha_c$, $\delta_c$) coordinates of the cluster centers, along with their redshifts ($z_c$). For comparison, Table~\ref{tab:pos} also shows the cluster redshifts presented in \citetalias{Biviano16} and \citetalias{Biviano21} which are in close agreement with this work. 
Any systematic differences in membership identification and/or dynamical properties between studies are, therefore, unlikely to be attributable to a different value of cluster redshift.
Figure~\ref{fig:PS} shows phase-space diagrams ($\mathrm{v_{pec}}$ versus $r_p$) for the spectroscopically confirmed galaxies shown in Figure~\ref{fig:Coors}.

Red points show spectroscopically-confirmed galaxies within the GalWeight cluster boundary (optimal countour black line) and within $R_{200}$  (vertical dashed line). 
Table~\ref{tab:dyn} shows $R_{200}$, within which we calculate $\mathrm{N_{200}}$ (number of spectroscopic members within $R_{200}$), $\sigma_{200}$, and $\mathrm{M}_{200}$.
The blue circles in Figure~\ref{fig:PS} indicate the cluster members which were identified by \citetalias{Biviano21}. In Table~\ref{tab:dyn} we also show 
the corresponding results from \citetalias{Biviano21}, except for $\mathrm{N_{200}}$ and $\sigma_{200}$, where we instead provide their line-of-sight velocity dispersion, $\sigma_{los}$, which was calculated using all member galaxies.

\begin{figure}
\hspace{-0cm}    \includegraphics[width=1\linewidth]{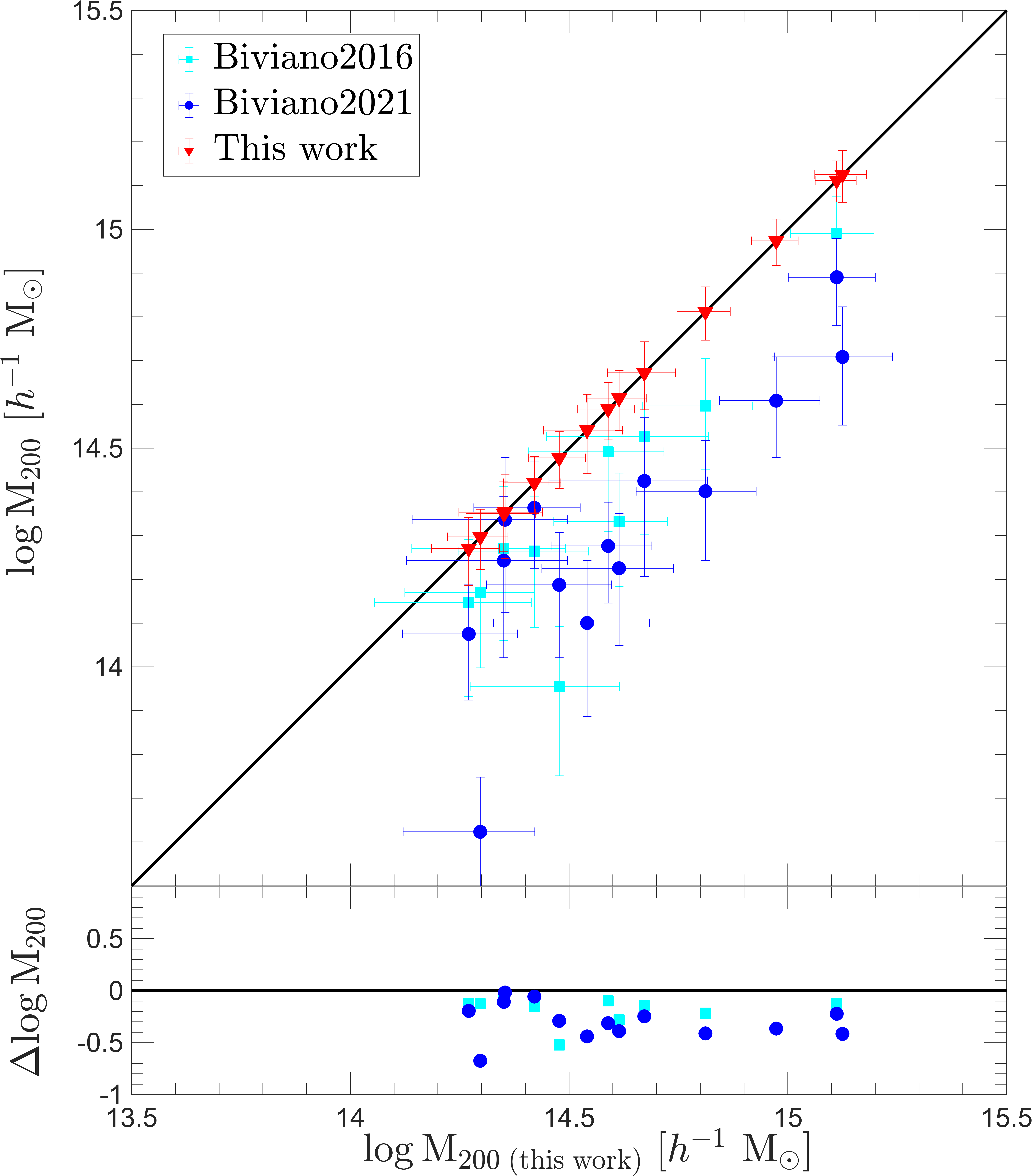} \vspace{-.5cm}
    \caption{A comparison that shows our results (the red points on the solid black line, representing 1:1 relation) in contrast with those from \citetalias{Biviano16} (cyan dots) and \citetalias{Biviano21} (blue dots) for $\mathrm{M}_{200}$. The bottom panel presents the scatter between the masses reported by \citetalias{Biviano16} and \citetalias{Biviano21} and our results, defined as $\Delta\log{\mathrm{M_{200}}}=\log{\mathrm{M_{200}}}(\mathrm{Biviano})-\log{\mathrm{M_{200}}}(\mathrm{ours})$.}
   \label{fig:comp}
\end{figure}

We observe from Figure \ref{fig:PS} that the members identified by GalWeight do not completely match those identified by \citetalias{Biviano21}. In the projected redshift space, the velocity distribution of galaxies in a cluster shows a characteristic trend: the velocity amplitude (or dispersion) is higher near the cluster center and decreases as the projected distance from the center increases (\citealp{Regos89,Praton94,Diaferio99,Abdullah13}). This behavior reflects the stronger gravitational binding in the inner regions, where galaxies move faster due to the cluster potential well, while galaxies in the outer regions have lower peculiar velocities, as they are less gravitationally bound. Consequently, the velocity distribution becomes narrower at larger projected distances, indicating a decline in the average peculiar velocity as we move away from the cluster center. This is also supported by the shape of the simulated galaxy clusters in redshift space (e.g., \citealp{Abdullah18,Paez22}). The caustic shape is clearly captured in simulations, as demonstrated by the application of the GalWeight technique to mock galaxy clusters (see \citealt{Abdullah18}). This is reflected in Figure \ref{fig:PS}, where the caustics identified by GalWeight generally perform well in separating cluster members from background galaxies.
It effectively shows a decline in velocity amplitude with increasing projected distance. This suggests that GalWeight provides a reliable approach for identifying cluster members, supported by its high efficiency in correctly identifying over 98\% of members in simulations \citep{Abdullah18}. Additionally, GalWeight has proven to be highly effective compared to four widely used cluster membership methods: the shifting gapper method, den Hartog method, the caustic technique, and the Spherical Infall Model \citep{Abdullah18}. 

Figure \ref{fig:comp} compares our estimated cluster mass $\mathrm{M}_{200}$ (the red points on the solid black line, representing 1:1 relation) in contrast with those from \citetalias{Biviano16} (cyan dots) and \citetalias{Biviano21} (blue dots) for $\mathrm{M}_{200}$. Notice that only 10 of the clusters in our sample (excluding SpARCS0335, SPT0205, SPT0546, and SPT2106) have been studied by \citetalias{Biviano16}. The bottom panel presents the scatter between the masses reported by \citetalias{Biviano16} and \citetalias{Biviano21} and our results, defined as $\Delta\log{\mathrm{M_{200}}}=\log{\mathrm{M_{200}}}(\mathrm{Biviano})-\log{\mathrm{M_{200}}}(\mathrm{ours})$.
This figure highlights the differences in cluster mass values obtained from our analysis using the virial mass estimator, in contrast to those derived from both \citetalias{Biviano16} and \citetalias{Biviano21}, who employed the $\mathtt{MAMPOSSt}$ method. It also stress the discrepancies between the two studies, \citetalias{Biviano16} and \citetalias{Biviano21}, despite their application of the same $\mathtt{MAMPOSSt}$ technique. 
The performance of the virial mass estimator and the $\mathtt{MAMPOSSt}$ method in recovering cluster mass was assessed using two mock catalogs, HOD2 and SAM2, as referenced in \citet{Old15}. They employed various statistics to examine the performance of the mass reconstruction methods. These statistics include the root-mean-square ($\mathrm{rms}$) difference between the recovered and true log mass, the scatter in the recovered mass ($\sigma_{\mathrm{M}_\mathrm{rec}}$), the scatter about the true mass ($\sigma_{\mathrm{M}_\mathrm{true}}$), and the bias at the pivot mass (taken as the median log mass of the input cluster sample) between the recovered and the true mass (see \citealp{Old15} and the references therein). The results indicated that the virial mass estimator outperforms the $\mathtt{MAMPOSSt}$ technique. Specifically, $\mathrm{rms}$, $\sigma_{\mathrm{M}_\mathrm{rec}}$, $\sigma_{\mathrm{M}_\mathrm{{true}}}$, and the bias values for the virial mass estimator returned \(0.24\), \(0.23\), \(0.23\), and \(0.06\) for HOD2, and \(0.32\), \(0.21\), \(0.23\), and \(0.24\) for SAM2, respectively \citep{Abdullah20a}. In comparison, for the $\mathtt{MAMPOSSt}$ method, these values were \(0.37\), \(0.29\), \(0.27\), and \(-0.19\) for HOD2, and \(0.31\), \(0.24\), \(0.32\), and \(-0.18\) for SAM2, respectively \citep{Old15}. 
We attribute the differences in cluster mass estimates between our results and those of \citetalias{Biviano16} and \citetalias{Biviano21} primarily to the choice of cluster mass estimator, and partially to the membership method — and the resulting slight differences in the measured velocity dispersions compared to those reported by \citetalias{Biviano21}.

\begin{table*}                                                                                                             
\centering                                                                                                                 
\begin{threeparttable}                                                                                                     
\caption{Fitting parameters for the velocity dispersion-cluster mass relation. }            
\label{tab:fit}                                                                                                            
\scriptsize                                                                                                                
\begin{tabular}{cccc} 
\toprule                                        
\multicolumn{4}{c}{Velocity Dispersion-Mass Relation ($\sigma$MR) at $\mathrm{M}_{\mathrm{piv}}=0.5\times10^{15}$ \hm} \\  
\midrule                                         &Normalization ($\alpha$) [\ks] &  Slope ($\beta$)  &   Intrinsic Uncertainty ($\sigma_{\mathrm{int}}$) \\
\midrule                                                                         
This Work  & 2.940 $\pm$ 0.028   & 0.373 $\pm$ 0.089    & 0.023 $\pm$ 0.024 \\
\citetalias{Biviano21}\tnote{$\ast$} & 3.000 $\pm$ 0.025  & 0.351 $\pm$ 0.051  & 0.015 $\pm$ 0.017\\ 
\citet{Evrard08}& $2.933\pm0.002$ & $0.336\pm0.003$ & $0.043\pm0.015$\\
\citet{Munari13}-DM& $2.939\pm0.002$ & $0.334\pm0.001$ & ---\\
\citet{Munari13}-CSF& $2.960\pm0.003$ & $0.355\pm0.003$ & ---\\
\citet{Armitage18}& $2.939\pm0.005$ & $0.350\pm0.010$ & ---\\
\bottomrule         
\end{tabular}
\begin{tablenotes}
    \item[$\ast$] For \citetalias{Biviano21}, $\sigma_{200}$ values were not explicitly provided in their paper. Instead, \citetalias{Biviano21} calculated the velocity dispersion ($\sigma_{los}$) using all member galaxies identified within the cluster field of view. In our analysis, we treat their $\sigma_{los}$ as $\sigma_{200}$ (since the two quantities are expected to be quite similar).
\end{tablenotes}
\end{threeparttable}
\end{table*}

\section{Velocity Dispersion-Cluster Mass Relation}
\label{sec:richness}
In this section, we introduce our methodology for fitting the scaling relation of $\sigma$MR. Then, we apply it to the GG sample to derive best-fit parameters of normalization $\alpha$, slope $\beta$, and intrinsic scatter $\sigma_\mathrm{int}$.

\begin{figure*}\hspace{-0cm}    \includegraphics[width=1\linewidth]{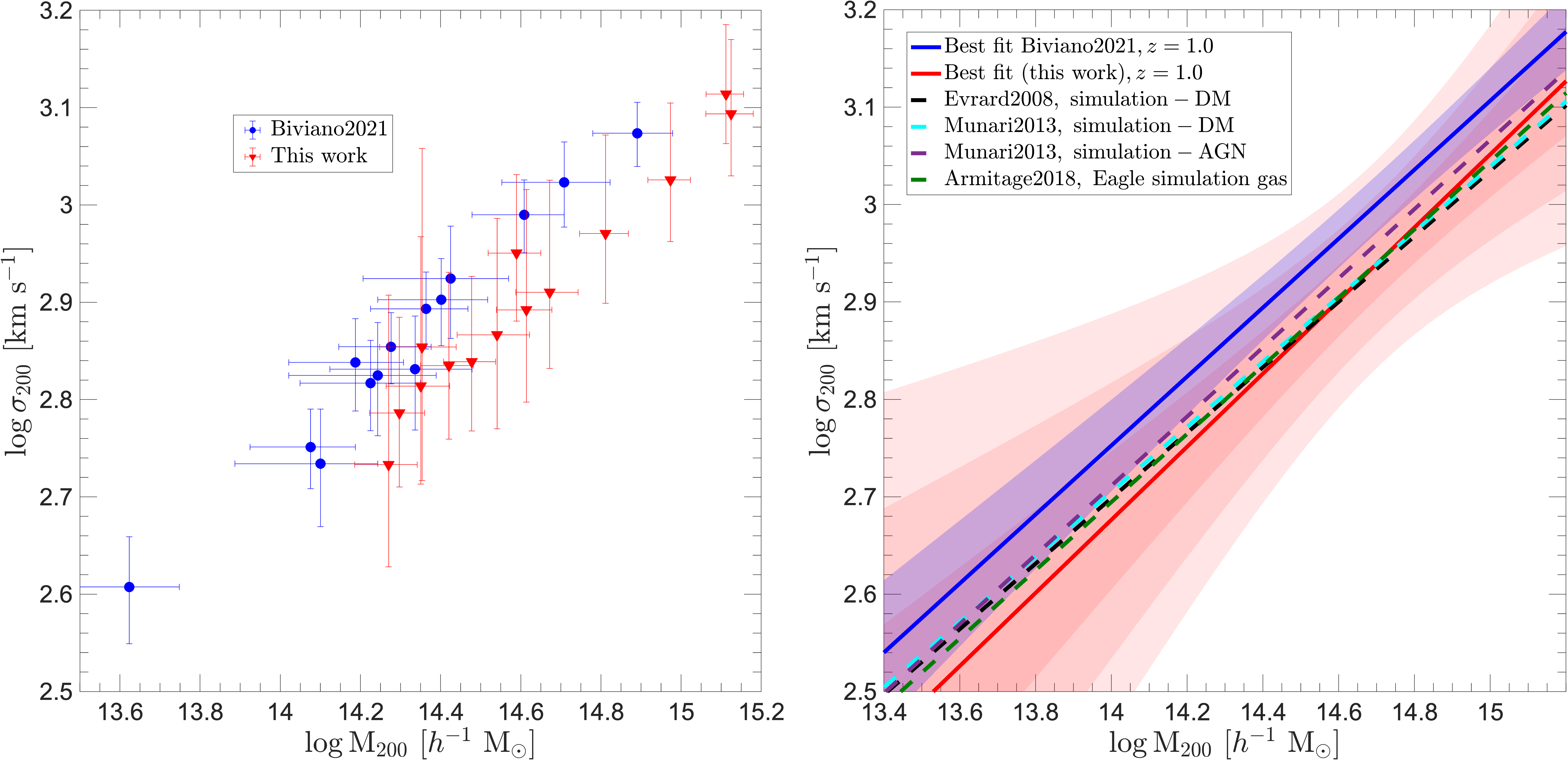} \vspace{-.5cm}
    \caption{Velocity dispersion $\sigma_{200}$ vs. virial mass $\mathrm{M}_{200}$. Left: the distribution of the cluster sample in the $\sigma_{200}$ - $\mathrm{M}_{200}$ plane, where red and blue points represent the results from our work and \citetalias{Biviano21}, respectively. Note that for \citetalias{Biviano21}, the $\sigma$ values correspond to $\sigma_{los}$ (the velocity dispersion for all members), since they did not provide $\sigma_{200}$ values. Right: the red and blue solid lines represent the best fit at $z=1$ for our work and \citetalias{Biviano21}, respectively. For comparison, we include relations from $N$-body dark matter simulations, specifically those derived by \citet{Evrard08} (dashed black) and \citet{Munari13} (dashed cyan), as well as from simulations incorporating baryonic effects, including \citet{Munari13}-AGN (dashed violet) and \citet{Armitage18}-Eagle simulation (dashed green). The three shaded red areas indicate $1\sigma$, $2\sigma$, and $3\sigma$ uncertainties for our best-fit line and the blue shaded area is $1\sigma$ for \citetalias{Biviano21}}.
   \label{fig:SigM}
\end{figure*}

\subsection{Methodology of fitting the Scaling Relations} \label{sec:method} 
The probability distribution of a dependent variable Y with a fixed independent variable X is given by a lognormal distribution (e.g., \citealp{Saro15,Simet17,Chiu20}) as
\begin{equation} \label{eq:prob}
\begin{split}
P(\log{Y}|X)= \frac{1}{\sqrt{2\pi\sigma^2_{\log{Y},X}}} \times ~~~~~~~~~~~~~~~~~~~~~\\
\exp{\left[-\frac{\left(\log{Y} - \left<\log{Y}|X\right>\right)^2}{2\sigma^2_{\log{Y},X}}\right]},
\end{split}
\end{equation}
\noindent where 
\begin{equation} \label{eq:rich}
\left<\log{Y}|X\right> = \alpha +\beta \log{X},
\end{equation}
In addition, the total variance in Y ($\sigma^2_{\log{Y},X}$) at a fixed X, including contributions of the X measurement errors ($\sigma_{\log{X}}$), Y measurement errors ($\sigma_{\log{Y}}$), and the intrinsic scatter in Y ($\sigma_\mathrm{int}$), is modeled by
\begin{equation} \label{eq:var}
\sigma^2_{\log{Y},X} = \beta^2\sigma^2_{\log{X}}+\sigma^2_{\log{Y}}+\sigma^2_{int}
\end{equation}
\noindent where $\alpha$ is the normalization and $\beta$ is the slope of the Y-X relation. 
The intrinsic scatter, $\sigma_{\mathrm{int}}$, quantifies the natural variance in $Y$ at fixed $X$ that is not accounted for by measurement uncertainties in either $X$ or $Y$. In our analysis, $\sigma_{\mathrm{int}}$ is treated as a constant free parameter and is fitted simultaneously with $\alpha$ and $\beta$.

We estimate the model parameters $\alpha, \beta$, and $\sigma_\mathrm{int}$ with the affine-invariant Markov chain Monte Carlo (hereafter MCMC) sampler of \citet{Goodman10} as implemented in the MATLAB package GWMCMC \footnote{\url{https://github.com/grinsted/gwmcmc}} inspired by the python package $\mathtt{emcee}$ \citep{Foreman13}.

To fit for the $\sigma$MR relation, we assume $Y\equiv \sigma_{200}$ and $X\equiv \mathrm{M_{200}}$.  The best-fit parameters for the GG sample at $\mathrm{M}_{\mathrm{piv}}=5\times10^{14}$ \hm~are presented in Table \ref{tab:fit}. We also carry out the $\sigma$MR fit on the data obtained from \citetalias{Biviano21}, as listed in the table. Notably, the values of $\sigma_{200}$ were not explicitly provided in \citetalias{Biviano21}. Therefore, we used the available and reported line-of-sight velocity dispersion $\sigma_{los}$ values (calculated for all members), as the two quantities are expected to be similar. 
Figure \ref{fig:SigM} shows the velocity dispersion $\sigma_{200}$ versus virial mass $\mathrm{M}_{200}$. The left panel displays the distribution of the cluster sample in the $\sigma_{200}$ - $\mathrm{M}_{200}$ plane. The red and blue points show the results for our work and \citetalias{Biviano21}, respectively. The right panel shows the best fit for each sample, with the red and blue solid lines representing the best fits at $z=1$ for our work and \citetalias{Biviano21}, respectively. The shaded areas indicate $1\sigma$, $2\sigma$, and $3\sigma$ uncertainties. To provide a broader context for our results, we compare them to theoretical relations derived from $N$-body dark matter simulations, such as those presented by \citet{Evrard08} and \citet{Munari13}, represented by dashed black and dashed cyan lines, respectively. Additionally, we include comparisons to simulations that account for baryonic effects, specifically \citet{Munari13}-AGN (dashed violet) and \citet{Armitage18}-Eagle simulation (dashed yellow). 

These comparisons highlight the agreement between our results and theoretical results across different simulation approaches. For \citetalias{Biviano21}, the best fit relation exhibits a higher normalization while maintaining nearly the same slope as the simulations. We attribute this offset primarily to the lower mass estimates reported by \citetalias{Biviano21} relative to the simulation-based relation. In particular, Figure~\ref{fig:comp} illustrates that \citetalias{Biviano21} mass estimates are systematically lower than ours, which is why our results exhibit better agreement with the simulations. The slight differences in the calculated velocity dispersion values also play a minor role. Quantitatively, the quantity $\sigma_\mathrm{log}=\sqrt{\sum{(\log{\sigma}_\mathrm{200,samp}-\log{\sigma}_\mathrm{200,Evr}})^2/(N-1)}$ between the \citet{Evrard08} relation and each sample returns 0.029 and 0.065 for our analysis and \citetalias{Biviano21}, respectively.

\section{Conclusion} \label{sec:conc}
In this paper, we reanalyzed the GOGREEN/GCLASS (GG) cluster sample, consisting of 14 clusters. We employed the two-dimensional Adaptive Kernel Method (2DAKM) to identify the centers of these clusters, with the coordinates and redshifts provided in Table \ref{tab:pos}. Our findings show strong agreement with the values reported by \citetalias{Biviano16} and \citetalias{Biviano21}, confirming that all studies are based on almost the same cluster centers. This consistency ensures that the phase-space diagrams constructed for each cluster across the three studies are nearly identical, indicating that any differences in membership identification or dynamical properties are not due to center offsets. 

We then utilized the GalWeight technique to determine cluster members. Our findings, shown in Figure \ref{fig:PS}, reveal that the membership identified by GalWeight does not completely match that of \citetalias{Biviano21}; however, the caustic shape is accurately represented using the GalWeight method. Subsequently, we calculated the cluster mass \( \mathrm{M_{200}} \) using the virial mass estimator and determined the number of spectroscopic members \( \mathrm{N_{200}} \), as well as their velocity dispersion \( \sigma_{200} \). Our analysis, presented in Figure \ref{fig:comp}, highlights notable differences in cluster mass values compared to those derived using the $\mathtt{MAMPOSSt}$ method in \citetalias{Biviano16} and \citetalias{Biviano21}. Despite applying the same dynamical technique, discrepancies between \citetalias{Biviano16} and \citetalias{Biviano21} are evident. 

The velocity dispersion-cluster mass relation ($\sigma$MR) further validates the effectiveness of our procedure, as our results closely match those obtained from the simulations. In contrast, the fit derived from \citetalias{Biviano21} data (see Figure \ref{fig:SigM}) exhibits less agreement with the simulations.

In summary, our findings confirm the effectiveness of the GalWeight technique and the virial mass estimator in determining cluster membership and dynamical properties, respectively. Additionally, our findings show that velocity dispersion can be used directly to estimate cluster mass without relying on dynamical mass estimators.

\section*{Acknowledgements}
MA thanks the Instituto de Astrof\'{\i}sica de Andaluc\'{\i}a (IAA-CSIC), Centro de Supercomputaci\'on de Galicia (CESGA), and the Spanish academic and research network (RedIRIS) in Spain for hosting Uchuu DR1 in the Skies \& Universes site (\url{http://www.skiesanduniverses.org/}) for cosmological simulations.  The Uchuu simulations were carried
out on the Aterui II supercomputer at the Center for
Computational Astrophysics, CfCA, of the National Astronomical Observatory of Japan, and the K computer at the
RIKEN Advanced Institute for Computational Science. GW gratefully acknowledges support from the National Science Foundation through grant AST-2205189.

\setcounter{section}{0}
\renewcommand{\thesection}{Appendix~\Alph{section}}

\renewcommand{\thefigure}{A}

\section{Testing the Impact of Spectroscopic Incompleteness} \label{sec:incomp}

\begin{figure*}
\hspace{-0cm}    \includegraphics[width=1\linewidth]{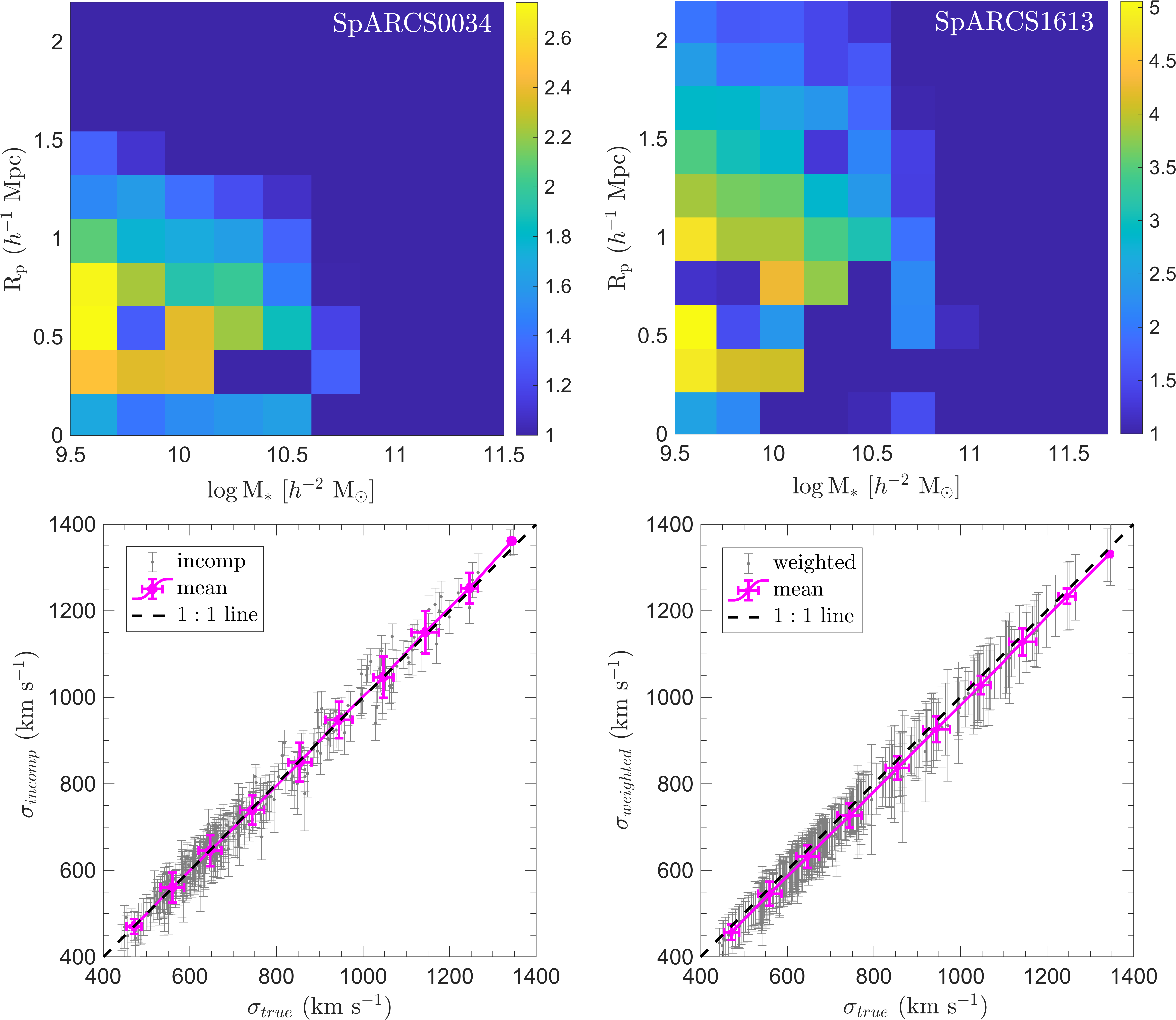} \vspace{-.5cm}
    \caption{Spectroscopic incompleteness test for two representative clusters: SpARCS0034 and SpARCS1613. Top panels show the weight functions in projected distance–stellar mass plane ($r_p$–$\log \mathrm{M}_\ast$), derived as the inverse of the completeness function (Section~\ref{sec:incomp}). Bottom panels compare the measured velocity dispersion from the incomplete sample ($\sigma_{\mathrm{incomp}}$, bottom left) and the weighted-corrected sample ($\sigma_{\mathrm{weighted}}$, bottom right) against the true dispersion ($\sigma_{\mathrm{true}}$) from fully sampled simulated clusters. Gray points represent the mean values across 500 Monte Carlo realizations, each randomly sampled based on the completeness function. The error bars show the standard deviation across those trials. Magenta points show the binned mean of the gray points with $1\sigma$ error bar; the dashed black line marks the one-to-one relation. While $\sigma_{\mathrm{incomp}}$ remains consistent with the true value, $\sigma_{\mathrm{weighted}}$ shows a slight systematic bias.}
   \label{fig:incomp}
\end{figure*}

Before calculating the dynamical properties of our cluster sample, we assess the spectroscopic incompleteness of the GG sample. 
Our approach models the incompleteness function individually for each cluster using the Uchuu-UM simulation. For each observed cluster, we estimate its halo mass using the incomplete spectroscopic data. We then select approximately 200 simulated clusters from Uchuu-UM with similar masses. From these simulated clusters, we include all galaxies with stellar masses above $\log \mathrm{M}_\ast = 9.5$~[\hmm]. By comparing the two-dimensional distribution of galaxies in projected distance–stellar mass space ($r_p$–$\log \mathrm{M}_\ast$) between the incomplete observed spectroscopic cluster and the stacked simulated clusters, we compute the completeness function as the ratio of the number density in the observed sample to that in the simulation. The weight function, defined as the inverse of the completeness function, is then used to statistically account for missing galaxies as a function of $r_p$ and $\log \mathrm{M}_\ast$.

To evaluate whether spectroscopic incompleteness biases the measurement of dynamical properties, we perform a controlled test using the Uchuu-UM simulation. We select a fully sampled simulated cluster with similar halo mass as the observed one and compute its line-of-sight velocity dispersion, $\sigma_{\mathrm{true}}$. The completeness function is applied to the simulated cluster to emulate the observational selection, from which we compute the velocity dispersion of the downsampled cluster, $\sigma_{\mathrm{incomp}}$. We then apply the weight function to the same incomplete simulated cluster to obtain a completeness-corrected estimate, $\sigma_{\mathrm{weighted}}$. This procedure is repeated 500 times by randomly sampling galaxies according to the completeness function in each realization. In each trial, we compute both $\sigma_{\mathrm{incomp}}$ and $\sigma_{\mathrm{weighted}}$, and we report their mean values and standard deviations. By comparing $\sigma_{\mathrm{true}}$, $\sigma_{\mathrm{incomp}}$, and $\sigma_{\mathrm{weighted}}$, we assess the impact of spectroscopic incompleteness on the velocity dispersion measurement, and whether applying a statistical correction is necessary.

Figure~\ref{fig:incomp} demonstrates our test on two representative clusters: SpARCS0034 and SpARCS1613. The upper panels show the spectroscopic weight functions derived for each cluster in the $r_p$–$\log \mathrm{M}_\ast$ plane. The lower-left panel compares the velocity dispersion measured from the incomplete realizations of simulated clusters, $\sigma_{\mathrm{incomp}}$, to the true dispersion values, $\sigma_{\mathrm{true}}$. The lower-right panel shows the result after applying the weight function, yielding the corrected estimates $\sigma_{\mathrm{weighted}}$. In both panels, the gray points represent the mean velocity dispersion measured from 500 Monte Carlo realizations for individual clusters, with error bars showing the standard deviation across those realizations. The dashed black line represents the one-to-one relation, while the magenta points show the binned mean of the gray points with $1\sigma$ error bar.
We find that $\sigma_{\mathrm{incomp}}$ (left panel) is consistent with the one-to-one relation, indicating that the spectroscopic incompleteness does not significantly bias the velocity dispersion. However, the weighted estimates $\sigma_{\mathrm{weighted}}$ (right panel) show a slight systematic bias below the one-to-one line. This suggests that applying the weight correction may not improve—and could slightly degrade—the velocity dispersion estimate in this regime. Therefore, we proceed to calculate the dynamical properties of our cluster sample using the incomplete spectroscopic data without applying a weight-based correction. This is consistent with the argument presented by \citet{Biviano21}, who note that spectroscopic incompleteness is not expected to significantly affect the observed velocity distribution of cluster galaxies \citep{Biviano02, Adami98, Goto05}. They further emphasize that spectroscopic incompleteness becomes critical primarily in Jeans analyses, where the galaxy number density profile enters through its radial derivative. Since our analysis is based solely on velocity dispersion measurements and does not involve the Jeans equation or number density modeling, the radial variation in spectroscopic completeness does not impact our results.

\bibliography{Ref}{}

\begin{thebibliography}{}
\expandafter\ifx\csname natexlab\endcsname\relax\def\natexlab#1{#1}\fi
\providecommand{\url}[1]{\href{#1}{#1}}
\providecommand{\dodoi}[1]{doi:~\href{http://doi.org/#1}{\nolinkurl{#1}}}
\providecommand{\doeprint}[1]{\href{http://ascl.net/#1}{\nolinkurl{http://ascl.net/#1}}}
\providecommand{\doarXiv}[1]{\href{https://arxiv.org/abs/#1}{\nolinkurl{https://arxiv.org/abs/#1}}}

\bibitem[{{Abdullah} {et~al.}(2011){Abdullah}, {Ali}, {Ismail}, \& {Rassem}}]{Abdullah11}
{Abdullah}, M.~H., {Ali}, G.~B., {Ismail}, H.~A., \& {Rassem}, M.~A. 2011, \mnras, 416, 2027, \dodoi{10.1111/j.1365-2966.2011.19178.x}

\bibitem[{{Abdullah} {et~al.}(2024){Abdullah}, {Klypin}, {Prada}, {Wilson}, {Ishiyama}, \& {Ereza}}]{Abdullah24}
{Abdullah}, M.~H., {Klypin}, A., {Prada}, F., {et~al.} 2024, \mnras, 529, L54, \dodoi{10.1093/mnrasl/slad200}

\bibitem[{{Abdullah} {et~al.}(2013){Abdullah}, {Praton}, \& {Ali}}]{Abdullah13}
{Abdullah}, M.~H., {Praton}, E.~A., \& {Ali}, G.~B. 2013, \mnras, 434, 1989, \dodoi{10.1093/mnras/stt1145}

\bibitem[{{Abdullah} {et~al.}(2018){Abdullah}, {Wilson}, \& {Klypin}}]{Abdullah18}
{Abdullah}, M.~H., {Wilson}, G., \& {Klypin}, A. 2018, \apj, 861, 22, \dodoi{10.3847/1538-4357/aac5db}

\bibitem[{{Abdullah} {et~al.}(2020){Abdullah}, {Wilson}, {Klypin}, {Old}, {Praton}, \& {Ali}}]{Abdullah20a}
{Abdullah}, M.~H., {Wilson}, G., {Klypin}, A., {et~al.} 2020, \apjs, 246, 2, \dodoi{10.3847/1538-4365/ab536e}

\bibitem[{{Adami} {et~al.}(1998){Adami}, {Biviano}, \& {Mazure}}]{Adami98}
{Adami}, C., {Biviano}, A., \& {Mazure}, A. 1998, \aap, 331, 439, \dodoi{10.48550/arXiv.astro-ph/9709268}

\bibitem[{{Allen} {et~al.}(2011){Allen}, {Evrard}, \& {Mantz}}]{Allen11}
{Allen}, S.~W., {Evrard}, A.~E., \& {Mantz}, A.~B. 2011, \araa, 49, 409, \dodoi{10.1146/annurev-astro-081710-102514}

\bibitem[{{Armitage} {et~al.}(2018){Armitage}, {Barnes}, {Kay}, {Bah{\'e}}, {Dalla Vecchia}, {Crain}, \& {Theuns}}]{Armitage18}
{Armitage}, T.~J., {Barnes}, D.~J., {Kay}, S.~T., {et~al.} 2018, \mnras, 474, 3746, \dodoi{10.1093/mnras/stx3020}

\bibitem[{{Ashman} {et~al.}(1994){Ashman}, {Bird}, \& {Zepf}}]{Ashman94}
{Ashman}, K.~M., {Bird}, C.~M., \& {Zepf}, S.~E. 1994, \aj, 108, 2348, \dodoi{10.1086/117248}

\bibitem[{{Aung} {et~al.}(2023){Aung}, {Nagai}, {Klypin}, {Behroozi}, {Abdullah}, {Ishiyama}, {Prada}, {P{\'e}rez}, {L{\'o}pez Cacheiro}, \& {Ruedas}}]{Aung23}
{Aung}, H., {Nagai}, D., {Klypin}, A., {et~al.} 2023, \mnras, 519, 1648, \dodoi{10.1093/mnras/stac3514}

\bibitem[{{Bahcall} \& {Tremaine}(1981)}]{Bahcall81}
{Bahcall}, J.~N., \& {Tremaine}, S. 1981, \apj, 244, 805, \dodoi{10.1086/158756}

\bibitem[{{Bahcall}(1988)}]{Bahcall88}
{Bahcall}, N.~A. 1988, \araa, 26, 631, \dodoi{10.1146/annurev.aa.26.090188.003215}

\bibitem[{Balogh {et~al.}(2017)Balogh, Gilbank, Muzzin, Rudnick, Cooper, Lidman, Biviano, Demarco, McGee, Nantais, Noble, Old, Wilson, Yee, Bellhouse, Cerulo, Chan, Pintos-Castro, Simpson, {van der Burg}, Zaritsky, Ziparo, Alonso, Bower, Lucia, Finoguenov, Lambas, Muriel, Parker, Rettura, Valotto, \& Wetzel}]{Balogh17}
Balogh, M., Gilbank, D., Muzzin, A., {et~al.} 2017, Monthly Notices of the Royal Astronomical Society, 470, 4168, \dodoi{10.1093/mnras/stx1370}

\bibitem[{{Balogh} {et~al.}(2021){Balogh}, {van der Burg}, {Muzzin}, {Rudnick}, {Wilson}, {Webb}, {Biviano}, {Boak}, {Cerulo}, {Chan}, {Cooper}, {Gilbank}, {Gwyn}, {Lidman}, {Matharu}, {McGee}, {Old}, {Pintos-Castro}, {Reeves}, {Shipley}, {Vulcani}, {Yee}, {Alonso}, {Bellhouse}, {Cooke}, {Davidson}, {De Lucia}, {Demarco}, {Drakos}, {Fillingham}, {Finoguenov}, {Forrest}, {Golledge}, {Jablonka}, {Lambas Garcia}, {McNab}, {Muriel}, {Nantais}, {Noble}, {Parker}, {Petter}, {Poggianti}, {Townsend}, {Valotto}, {Webb}, \& {Zaritsky}}]{Balogh21}
{Balogh}, M.~L., {van der Burg}, R. F.~J., {Muzzin}, A., {et~al.} 2021, \mnras, 500, 358, \dodoi{10.1093/mnras/staa3008}

\bibitem[{{Bartelmann}(1996)}]{Bartelmann96}
{Bartelmann}, M. 1996, \aap, 313, 697, \dodoi{10.48550/arXiv.astro-ph/9602053}

\bibitem[{{Battye} \& {Weller}(2003)}]{Battye03}
{Battye}, R.~A., \& {Weller}, J. 2003, \prd, 68, 083506, \dodoi{10.1103/PhysRevD.68.083506}

\bibitem[{{Beers} {et~al.}(1990){Beers}, {Flynn}, \& {Gebhardt}}]{Beers90}
{Beers}, T.~C., {Flynn}, K., \& {Gebhardt}, K. 1990, \aj, 100, 32, \dodoi{10.1086/115487}

\bibitem[{{Beers} {et~al.}(1991){Beers}, {Gebhardt}, {Forman}, {Huchra}, \& {Jones}}]{Beers91}
{Beers}, T.~C., {Gebhardt}, K., {Forman}, W., {Huchra}, J.~P., \& {Jones}, C. 1991, \aj, 102, 1581, \dodoi{10.1086/115982}

\bibitem[{{Behroozi} {et~al.}(2019){Behroozi}, {Wechsler}, {Hearin}, \& {Conroy}}]{Behroozi19}
{Behroozi}, P., {Wechsler}, R.~H., {Hearin}, A.~P., \& {Conroy}, C. 2019, \mnras, 488, 3143, \dodoi{10.1093/mnras/stz1182}

\bibitem[{{Behroozi} {et~al.}(2013{\natexlab{a}}){Behroozi}, {Wechsler}, \& {Wu}}]{Behroozi13a}
{Behroozi}, P.~S., {Wechsler}, R.~H., \& {Wu}, H.-Y. 2013{\natexlab{a}}, \apj, 762, 109, \dodoi{10.1088/0004-637X/762/2/109}

\bibitem[{{Behroozi} {et~al.}(2013{\natexlab{b}}){Behroozi}, {Wechsler}, {Wu}, {Busha}, {Klypin}, \& {Primack}}]{Behroozi13b}
{Behroozi}, P.~S., {Wechsler}, R.~H., {Wu}, H.-Y., {et~al.} 2013{\natexlab{b}}, \apj, 763, 18, \dodoi{10.1088/0004-637X/763/1/18}

\bibitem[{{Binney} \& {Tremaine}(1987)}]{Binney87}
{Binney}, J., \& {Tremaine}, S. 1987, {Galactic dynamics}

\bibitem[{{Biviano} {et~al.}(2002){Biviano}, {Katgert}, {Thomas}, \& {Adami}}]{Biviano02}
{Biviano}, A., {Katgert}, P., {Thomas}, T., \& {Adami}, C. 2002, \aap, 387, 8, \dodoi{10.1051/0004-6361:20020340}

\bibitem[{{Biviano} {et~al.}(2016){Biviano}, {van der Burg}, {Muzzin}, {Sartoris}, {Wilson}, \& {Yee}}]{Biviano16}
{Biviano}, A., {van der Burg}, R.~F.~J., {Muzzin}, A., {et~al.} 2016, \aap, 594, A51, \dodoi{10.1051/0004-6361/201628697}

\bibitem[{{Biviano} {et~al.}(2021){Biviano}, {van der Burg}, {Balogh}, {Munari}, {Cooper}, {De Lucia}, {Demarco}, {Jablonka}, {Muzzin}, {Nantais}, {Old}, {Rudnick}, {Vulcani}, {Wilson}, {Yee}, {Zaritsky}, {Cerulo}, {Chan}, {Finoguenov}, {Gilbank}, {Lidman}, {Pintos-Castro}, \& {Shipley}}]{Biviano21}
{Biviano}, A., {van der Burg}, R.~F.~J., {Balogh}, M.~L., {et~al.} 2021, \aap, 650, A105, \dodoi{10.1051/0004-6361/202140564}

\bibitem[{{Brodwin} {et~al.}(2010){Brodwin}, {Ruel}, {Ade}, {Aird}, {Andersson}, {Ashby}, {Bautz}, {Bazin}, {Benson}, {Bleem}, {Carlstrom}, {Chang}, {Crawford}, {Crites}, {de Haan}, {Desai}, {Dobbs}, {Dudley}, {Fazio}, {Foley}, {Forman}, {Garmire}, {George}, {Gladders}, {Gonzalez}, {Halverson}, {High}, {Holder}, {Holzapfel}, {Hrubes}, {Jones}, {Joy}, {Keisler}, {Knox}, {Lee}, {Leitch}, {Lueker}, {Marrone}, {McMahon}, {Mehl}, {Meyer}, {Mohr}, {Montroy}, {Murray}, {Padin}, {Plagge}, {Pryke}, {Reichardt}, {Rest}, {Ruhl}, {Schaffer}, {Shaw}, {Shirokoff}, {Song}, {Spieler}, {Stalder}, {Stanford}, {Staniszewski}, {Stark}, {Stubbs}, {Vanderlinde}, {Vieira}, {Vikhlinin}, {Williamson}, {Yang}, {Zahn}, \& {Zenteno}}]{Brodwin10}
{Brodwin}, M., {Ruel}, J., {Ade}, P.~A.~R., {et~al.} 2010, \apj, 721, 90, \dodoi{10.1088/0004-637X/721/1/90}

\bibitem[{{Carlberg} {et~al.}(1997){Carlberg}, {Yee}, \& {Ellingson}}]{Carlberg97}
{Carlberg}, R.~G., {Yee}, H.~K.~C., \& {Ellingson}, E. 1997, \apj, 478, 462, \dodoi{10.1086/303805}

\bibitem[{{Carlberg} {et~al.}(1996){Carlberg}, {Yee}, {Ellingson}, {Abraham}, {Gravel}, {Morris}, \& {Pritchet}}]{Carlberg96}
{Carlberg}, R.~G., {Yee}, H.~K.~C., {Ellingson}, E., {et~al.} 1996, \apj, 462, 32, \dodoi{10.1086/177125}

\bibitem[{{Chiu} {et~al.}(2020){Chiu}, {Umetsu}, {Murata}, {Medezinski}, \& {Oguri}}]{Chiu20}
{Chiu}, I.~N., {Umetsu}, K., {Murata}, R., {Medezinski}, E., \& {Oguri}, M. 2020, \mnras, 495, 428, \dodoi{10.1093/mnras/staa1158}

\bibitem[{{Danese} {et~al.}(1980){Danese}, {de Zotti}, \& {di Tullio}}]{Danese80}
{Danese}, L., {de Zotti}, G., \& {di Tullio}, G. 1980, \aap, 82, 322

\bibitem[{{den Hartog} \& {Katgert}(1996)}]{denHartog96}
{den Hartog}, R., \& {Katgert}, P. 1996, \mnras, 279, 349, \dodoi{10.1093/mnras/279.2.349}

\bibitem[{{Diaferio}(1999)}]{Diaferio99}
{Diaferio}, A. 1999, \mnras, 309, 610, \dodoi{10.1046/j 1365-8711.1999.02864.x}

\bibitem[{{Dong-P{\'a}ez} {et~al.}(2022){Dong-P{\'a}ez}, {Smith}, {Szewciw}, {Ereza}, {Abdullah}, {Hern{\'a}ndez-Aguayo}, {Trusov}, {Prada}, {Klypin}, {Ishiyama}, {Berlind}, {Zarrouk}, {L{\'o}pez Cacheiro}, \& {Ruedas}}]{Paez22}
{Dong-P{\'a}ez}, C.~A., {Smith}, A., {Szewciw}, A.~O., {et~al.} 2022, arXiv e-prints, arXiv:2208.00540.
\newblock \doarXiv{2208.00540}

\bibitem[{{Evrard} {et~al.}(2008){Evrard}, {Bialek}, {Busha}, {White}, {Habib}, {Heitmann}, {Warren}, {Rasia}, {Tormen}, {Moscardini}, {Power}, {Jenkins}, {Gao}, {Frenk}, {Springel}, {White}, \& {Diemand}}]{Evrard08}
{Evrard}, A.~E., {Bialek}, J., {Busha}, M., {et~al.} 2008, \apj, 672, 122, \dodoi{10.1086/521616}

\bibitem[{{Fadda} {et~al.}(1996){Fadda}, {Girardi}, {Giuricin}, {Mardirossian}, \& {Mezzetti}}]{Fadda96}
{Fadda}, D., {Girardi}, M., {Giuricin}, G., {Mardirossian}, F., \& {Mezzetti}, M. 1996, \apj, 473, 670, \dodoi{10.1086/178180}

\bibitem[{{Finoguenov} {et~al.}(2007){Finoguenov}, {Guzzo}, {Hasinger}, {Scoville}, {Aussel}, {B{\"o}hringer}, {Brusa}, {Capak}, {Cappelluti}, {Comastri}, {Giodini}, {Griffiths}, {Impey}, {Koekemoer}, {Kneib}, {Leauthaud}, {Le F{\`e}vre}, {Lilly}, {Mainieri}, {Massey}, {McCracken}, {Mobasher}, {Murayama}, {Peacock}, {Sakelliou}, {Schinnerer}, {Silverman}, {Smol{\v{c}}i{\'c}}, {Taniguchi}, {Tasca}, {Taylor}, {Trump}, \& {Zamorani}}]{Finoguenov07}
{Finoguenov}, A., {Guzzo}, L., {Hasinger}, G., {et~al.} 2007, \apjs, 172, 182, \dodoi{10.1086/516577}

\bibitem[{{Finoguenov} {et~al.}(2010){Finoguenov}, {Watson}, {Tanaka}, {Simpson}, {Cirasuolo}, {Dunlop}, {Peacock}, {Farrah}, {Akiyama}, {Ueda}, {Smol{\v{c}}i{\'c}}, {Stewart}, {Rawlings}, {van Breukelen}, {Almaini}, {Clewley}, {Bonfield}, {Jarvis}, {Barr}, {Foucaud}, {McLure}, {Sekiguchi}, \& {Egami}}]{Finoguenov10}
{Finoguenov}, A., {Watson}, M.~G., {Tanaka}, M., {et~al.} 2010, \mnras, 403, 2063, \dodoi{10.1111/j.1365-2966.2010.16256.x}

\bibitem[{{Foley} {et~al.}(2011){Foley}, {Andersson}, {Bazin}, {de Haan}, {Ruel}, {Ade}, {Aird}, {Armstrong}, {Ashby}, {Bautz}, {Benson}, {Bleem}, {Bonamente}, {Brodwin}, {Carlstrom}, {Chang}, {Clocchiatti}, {Crawford}, {Crites}, {Desai}, {Dobbs}, {Dudley}, {Fazio}, {Forman}, {Garmire}, {George}, {Gladders}, {Gonzalez}, {Halverson}, {High}, {Holder}, {Holzapfel}, {Hoover}, {Hrubes}, {Jones}, {Joy}, {Keisler}, {Knox}, {Lee}, {Leitch}, {Lueker}, {Luong-Van}, {Marrone}, {McMahon}, {Mehl}, {Meyer}, {Mohr}, {Montroy}, {Murray}, {Padin}, {Plagge}, {Pryke}, {Reichardt}, {Rest}, {Ruhl}, {Saliwanchik}, {Saro}, {Schaffer}, {Shaw}, {Shirokoff}, {Song}, {Spieler}, {Stalder}, {Stanford}, {Staniszewski}, {Stark}, {Story}, {Stubbs}, {Vanderlinde}, {Vieira}, {Vikhlinin}, {Williamson}, \& {Zenteno}}]{Foley11}
{Foley}, R.~J., {Andersson}, K., {Bazin}, G., {et~al.} 2011, \apj, 731, 86, \dodoi{10.1088/0004-637X/731/2/86}

\bibitem[{{Foreman-Mackey} {et~al.}(2013){Foreman-Mackey}, {Hogg}, {Lang}, \& {Goodman}}]{Foreman13}
{Foreman-Mackey}, D., {Hogg}, D.~W., {Lang}, D., \& {Goodman}, J. 2013, \pasp, 125, 306, \dodoi{10.1086/670067}

\bibitem[{{Gao} {et~al.}(2008){Gao}, {Navarro}, {Cole}, {Frenk}, {White}, {Springel}, {Jenkins}, \& {Neto}}]{Gao08}
{Gao}, L., {Navarro}, J.~F., {Cole}, S., {et~al.} 2008, \mnras, 387, 536, \dodoi{10.1111/j.1365-2966.2008.13277.x}

\bibitem[{{George} {et~al.}(2011){George}, {Leauthaud}, {Bundy}, {Finoguenov}, {Tinker}, {Lin}, {Mei}, {Kneib}, {Aussel}, {Behroozi}, {Busha}, {Capak}, {Coccato}, {Covone}, {Faure}, {Fiorenza}, {Ilbert}, {Le Floc'h}, {Koekemoer}, {Tanaka}, {Wechsler}, \& {Wolk}}]{George11}
{George}, M.~R., {Leauthaud}, A., {Bundy}, K., {et~al.} 2011, \apj, 742, 125, \dodoi{10.1088/0004-637X/742/2/125}

\bibitem[{{Girardi} {et~al.}(1993){Girardi}, {Biviano}, {Giuricin}, {Mardirossian}, \& {Mezzetti}}]{Girardi93}
{Girardi}, M., {Biviano}, A., {Giuricin}, G., {Mardirossian}, F., \& {Mezzetti}, M. 1993, \apj, 404, 38, \dodoi{10.1086/172256}

\bibitem[{{Girardi} {et~al.}(1998){Girardi}, {Giuricin}, {Mardirossian}, {Mezzetti}, \& {Boschin}}]{Girardi98a}
{Girardi}, M., {Giuricin}, G., {Mardirossian}, F., {Mezzetti}, M., \& {Boschin}, W. 1998, \apj, 505, 74, \dodoi{10.1086/306157}

\bibitem[{{Golse} \& {Kneib}(2002)}]{Golse02}
{Golse}, G., \& {Kneib}, J.-P. 2002, \aap, 390, 821, \dodoi{10.1051/0004-6361:20020639}

\bibitem[{{Goodman} \& {Weare}(2010)}]{Goodman10}
{Goodman}, J., \& {Weare}, J. 2010, Communications in Applied Mathematics and Computational Science, 5, 65, \dodoi{10.2140/camcos.2010.5.65}

\bibitem[{{Goto}(2005)}]{Goto05}
{Goto}, T. 2005, \mnras, 359, 1415, \dodoi{10.1111/j.1365-2966.2005.08982.x}

\bibitem[{{Haiman} {et~al.}(2001){Haiman}, {Mohr}, \& {Holder}}]{Haiman01}
{Haiman}, Z., {Mohr}, J.~J., \& {Holder}, G.~P. 2001, \apj, 553, 545, \dodoi{10.1086/320939}

\bibitem[{{Heisler} {et~al.}(1985){Heisler}, {Tremaine}, \& {Bahcall}}]{Heisler85}
{Heisler}, J., {Tremaine}, S., \& {Bahcall}, J.~N. 1985, \apj, 298, 8, \dodoi{10.1086/163584}

\bibitem[{{Hook} {et~al.}(2004){Hook}, {J{\o}rgensen}, {Allington-Smith}, {Davies}, {Metcalfe}, {Murowinski}, \& {Crampton}}]{Hook04}
{Hook}, I.~M., {J{\o}rgensen}, I., {Allington-Smith}, J.~R., {et~al.} 2004, \pasp, 116, 425, \dodoi{10.1086/383624}

\bibitem[{{Ishiyama} {et~al.}(2009){Ishiyama}, {Fukushige}, \& {Makino}}]{Ishiyama09}
{Ishiyama}, T., {Fukushige}, T., \& {Makino}, J. 2009, \pasj, 61, 1319, \dodoi{10.1093/pasj/61.6.1319}

\bibitem[{{Ishiyama} {et~al.}(2012){Ishiyama}, {Nitadori}, \& {Makino}}]{Ishiyama12}
{Ishiyama}, T., {Nitadori}, K., \& {Makino}, J. 2012, arXiv e-prints, arXiv:1211.4406.
\newblock \doarXiv{1211.4406}

\bibitem[{{Ishiyama} {et~al.}(2021){Ishiyama}, {Prada}, {Klypin}, {Sinha}, {Metcalf}, {Jullo}, {Altieri}, {Cora}, {Croton}, {de la Torre}, {Mill{\'a}n-Calero}, {Oogi}, {Ruedas}, \& {Vega-Mart{\'\i}nez}}]{Ishiyama21}
{Ishiyama}, T., {Prada}, F., {Klypin}, A.~A., {et~al.} 2021, \mnras, 506, 4210, \dodoi{10.1093/mnras/stab1755}

\bibitem[{{Klypin} {et~al.}(2016){Klypin}, {Yepes}, {Gottl{\"o}ber}, {Prada}, \& {He{\ss}}}]{Klypin16}
{Klypin}, A., {Yepes}, G., {Gottl{\"o}ber}, S., {Prada}, F., \& {He{\ss}}, S. 2016, \mnras, 457, 4340, \dodoi{10.1093/mnras/stw248}

\bibitem[{{Koranyi} \& {Geller}(2000)}]{Koranyi00}
{Koranyi}, D.~M., \& {Geller}, M.~J. 2000, \aj, 119, 44, \dodoi{10.1086/301166}

\bibitem[{{Kravtsov} \& {Borgani}(2012)}]{Kravtsov12}
{Kravtsov}, A.~V., \& {Borgani}, S. 2012, \araa, 50, 353, \dodoi{10.1146/annurev-astro-081811-125502}

\bibitem[{{Limber} \& {Mathews}(1960)}]{Limber60}
{Limber}, D.~N., \& {Mathews}, W.~G. 1960, \apj, 132, 286, \dodoi{10.1086/146928}

\bibitem[{{Mamon} {et~al.}(2013){Mamon}, {Biviano}, \& {Bou{\'e}}}]{Mamon13}
{Mamon}, G.~A., {Biviano}, A., \& {Bou{\'e}}, G. 2013, \mnras, 429, 3079, \dodoi{10.1093/mnras/sts565}

\bibitem[{{Mamon} {et~al.}(2010){Mamon}, {Biviano}, \& {Murante}}]{Mamon10}
{Mamon}, G.~A., {Biviano}, A., \& {Murante}, G. 2010, \aap, 520, A30, \dodoi{10.1051/0004-6361/200913948}

\bibitem[{{Mamon} \& {Bou{\'e}}(2010)}]{Mamon&Boue10}
{Mamon}, G.~A., \& {Bou{\'e}}, G. 2010, \mnras, 401, 2433, \dodoi{10.1111/j.1365-2966.2009.15817.x}

\bibitem[{{Mamon} \& {{\L}okas}(2005)}]{Mamon&Lokas05}
{Mamon}, G.~A., \& {{\L}okas}, E.~L. 2005, \mnras, 363, 705, \dodoi{10.1111/j.1365-2966.2005.09400.x}

\bibitem[{{McConachie} {et~al.}(2022){McConachie}, {Wilson}, {Forrest}, {Marsan}, {Muzzin}, {Cooper}, {Annunziatella}, {Marchesini}, {Chan}, {Gomez}, {Abdullah}, {Saracco}, \& {Nantais}}]{McConachie22}
{McConachie}, I., {Wilson}, G., {Forrest}, B., {et~al.} 2022, \apj, 926, 37, \dodoi{10.3847/1538-4357/ac2b9f}

\bibitem[{Mclachlan \& Basford(1988)}]{Mclachlan88}
Mclachlan, G., \& Basford, K. 1988, Mixture Models: Inference and Applications to Clustering, Vol.~38, \dodoi{10.2307/2348072}

\bibitem[{{Munari} {et~al.}(2013){Munari}, {Biviano}, {Borgani}, {Murante}, \& {Fabjan}}]{Munari13}
{Munari}, E., {Biviano}, A., {Borgani}, S., {Murante}, G., \& {Fabjan}, D. 2013, \mnras, 430, 2638, \dodoi{10.1093/mnras/stt049}

\bibitem[{{Muzzin} {et~al.}(2009){Muzzin}, {Wilson}, {Yee}, {Hoekstra}, {Gilbank}, {Surace}, {Lacy}, {Blindert}, {Majumdar}, {Demarco}, {Gardner}, {Gladders}, \& {Lonsdale}}]{Muzzin09}
{Muzzin}, A., {Wilson}, G., {Yee}, H.~K.~C., {et~al.} 2009, \apj, 698, 1934, \dodoi{10.1088/0004-637X/698/2/1934}

\bibitem[{{Muzzin} {et~al.}(2012){Muzzin}, {Wilson}, {Yee}, {Gilbank}, {Hoekstra}, {Demarco}, {Balogh}, {van Dokkum}, {Franx}, {Ellingson}, {Hicks}, {Nantais}, {Noble}, {Lacy}, {Lidman}, {Rettura}, {Surace}, \& {Webb}}]{Muzzin12}
---. 2012, \apj, 746, 188, \dodoi{10.1088/0004-637X/746/2/188}

\bibitem[{{Nantais} {et~al.}(2016){Nantais}, {van der Burg}, {Lidman}, {Demarco}, {Noble}, {Wilson}, {Muzzin}, {Foltz}, {DeGroot}, \& {Cooper}}]{Nantais16}
{Nantais}, J.~B., {van der Burg}, R. F.~J., {Lidman}, C., {et~al.} 2016, \aap, 592, A161, \dodoi{10.1051/0004-6361/201628663}

\bibitem[{{Navarro} {et~al.}(1996){Navarro}, {Frenk}, \& {White}}]{NFW96}
{Navarro}, J.~F., {Frenk}, C.~S., \& {White}, S.~D.~M. 1996, \apj, 462, 563, \dodoi{10.1086/177173}

\bibitem[{{Navarro} {et~al.}(1997){Navarro}, {Frenk}, \& {White}}]{NFW97}
---. 1997, \apj, 490, 493, \dodoi{10.1086/304888}

\bibitem[{{Nelson} {et~al.}(2018){Nelson}, {Springel}, {Pillepich}, {Rodriguez-Gomez}, {Torrey}, {Genel}, {Vogelsberger}, {Pakmor}, {Marinacci}, {Weinberger}, {Kelley}, {Lovell}, {Diemer}, \& {Hernquist}}]{Nelson18}
{Nelson}, D., {Springel}, V., {Pillepich}, A., {et~al.} 2018, arXiv e-prints.
\newblock \doarXiv{1812.05609}

\bibitem[{{Old} {et~al.}(2015){Old}, {Wojtak}, {Mamon}, {Skibba}, {Pearce}, {Croton}, {Bamford}, {Behroozi}, {de Carvalho}, {Mu{\~n}oz-Cuartas}, {Gifford}, {Gray}, {von der Linden}, {Merrifield}, {Muldrew}, {M{\"u}ller}, {Pearson}, {Ponman}, {Rozo}, {Rykoff}, {Saro}, {Sepp}, {Sif{\'o}n}, \& {Tempel}}]{Old15}
{Old}, L., {Wojtak}, R., {Mamon}, G.~A., {et~al.} 2015, \mnras, 449, 1897, \dodoi{10.1093/mnras/stv421}

\bibitem[{{Overzier}(2016)}]{Overzier16}
{Overzier}, R.~A. 2016, \aapr, 24, 14, \dodoi{10.1007/s00159-016-0100-3}

\bibitem[{{Perea} {et~al.}(1990){Perea}, {del Olmo}, \& {Moles}}]{Perea90a}
{Perea}, J., {del Olmo}, A., \& {Moles}, M. 1990, \aap, 237, 319

\bibitem[{{Pisani}(1993)}]{Pisani93}
{Pisani}, A. 1993, \mnras, 265, 706, \dodoi{10.1093/mnras/265.3.706}

\bibitem[{{Pisani}(1996)}]{Pisani96}
---. 1996, \mnras, 278, 697, \dodoi{10.1093/mnras/278.3.697}

\bibitem[{{Postman} {et~al.}(1992){Postman}, {Huchra}, \& {Geller}}]{Postman92}
{Postman}, M., {Huchra}, J.~P., \& {Geller}, M.~J. 1992, \apj, 384, 404, \dodoi{10.1086/170883}

\bibitem[{{Praton} \& {Schneider}(1994)}]{Praton94}
{Praton}, E.~A., \& {Schneider}, S.~E. 1994, \apj, 422, 46, \dodoi{10.1086/173702}

\bibitem[{{Regos} \& {Geller}(1989)}]{Regos89}
{Regos}, E., \& {Geller}, M.~J. 1989, \aj, 98, 755, \dodoi{10.1086/115177}

\bibitem[{{Reiprich} \& {B{\"o}hringer}(2002)}]{Reiprich02}
{Reiprich}, T.~H., \& {B{\"o}hringer}, H. 2002, \apj, 567, 716, \dodoi{10.1086/338753}

\bibitem[{{Rines} {et~al.}(2003){Rines}, {Geller}, {Kurtz}, \& {Diaferio}}]{Rines03}
{Rines}, K., {Geller}, M.~J., {Kurtz}, M.~J., \& {Diaferio}, A. 2003, \aj, 126, 2152, \dodoi{10.1086/378599}

\bibitem[{{Rozo} {et~al.}(2010){Rozo}, {Wechsler}, {Rykoff}, {Annis}, {Becker}, {Evrard}, {Frieman}, {Hansen}, {Hao}, {Johnston}, {Koester}, {McKay}, {Sheldon}, \& {Weinberg}}]{Rozo10}
{Rozo}, E., {Wechsler}, R.~H., {Rykoff}, E.~S., {et~al.} 2010, \apj, 708, 645, \dodoi{10.1088/0004-637X/708/1/645}

\bibitem[{{Rykoff} {et~al.}(2016){Rykoff}, {Rozo}, {Hollowood}, {Bermeo-Hernandez}, {Jeltema}, {Mayers}, {Romer}, {Rooney}, {Saro}, {Vergara Cervantes}, {Wechsler}, {Wilcox}, {Abbott}, {Abdalla}, {Allam}, {Annis}, {Benoit-L{\'e}vy}, {Bernstein}, {Bertin}, {Brooks}, {Burke}, {Capozzi}, {Carnero Rosell}, {Carrasco Kind}, {Castander}, {Childress}, {Collins}, {Cunha}, {D'Andrea}, {da Costa}, {Davis}, {Desai}, {Diehl}, {Dietrich}, {Doel}, {Evrard}, {Finley}, {Flaugher}, {Fosalba}, {Frieman}, {Glazebrook}, {Goldstein}, {Gruen}, {Gruendl}, {Gutierrez}, {Hilton}, {Honscheid}, {Hoyle}, {James}, {Kay}, {Kuehn}, {Kuropatkin}, {Lahav}, {Lewis}, {Lidman}, {Lima}, {Maia}, {Mann}, {Marshall}, {Martini}, {Melchior}, {Miller}, {Miquel}, {Mohr}, {Nichol}, {Nord}, {Ogando}, {Plazas}, {Reil}, {Sahl{\'e}n}, {Sanchez}, {Santiago}, {Scarpine}, {Schubnell}, {Sevilla-Noarbe}, {Smith}, {Soares-Santos}, {Sobreira}, {Stott}, {Suchyta}, {Swanson}, {Tarle}, {Thomas}, {Tucker}, {Uddin}, {Viana}, {Vikram}, {Walker}, {Zhang}, \& {DES
  Collaboration}}]{Rykoff16}
{Rykoff}, E.~S., {Rozo}, E., {Hollowood}, D., {et~al.} 2016, \apjs, 224, 1, \dodoi{10.3847/0067-0049/224/1/1}

\bibitem[{{Saro} {et~al.}(2015){Saro}, {Bocquet}, {Rozo}, {Benson}, {Mohr}, {Rykoff}, {Soares-Santos}, {Bleem}, {Dodelson}, {Melchior}, {Sobreira}, {Upadhyay}, {Weller}, {Abbott}, {Abdalla}, {Allam}, {Armstrong}, {Banerji}, {Bauer}, {Bayliss}, {Benoit-L{\'e}vy}, {Bernstein}, {Bertin}, {Brodwin}, {Brooks}, {Buckley-Geer}, {Burke}, {Carlstrom}, {Capasso}, {Capozzi}, {Carnero Rosell}, {Carrasco Kind}, {Chiu}, {Covarrubias}, {Crawford}, {Crocce}, {D'Andrea}, {da Costa}, {DePoy}, {Desai}, {de Haan}, {Diehl}, {Dietrich}, {Doel}, {Cunha}, {Eifler}, {Evrard}, {Fausti Neto}, {Fernandez}, {Flaugher}, {Fosalba}, {Frieman}, {Gangkofner}, {Gaztanaga}, {Gerdes}, {Gruen}, {Gruendl}, {Gupta}, {Hennig}, {Holzapfel}, {Honscheid}, {Jain}, {James}, {Kuehn}, {Kuropatkin}, {Lahav}, {Li}, {Lin}, {Maia}, {March}, {Marshall}, {Martini}, {McDonald}, {Miller}, {Miquel}, {Nord}, {Ogando}, {Plazas}, {Reichardt}, {Romer}, {Roodman}, {Sako}, {Sanchez}, {Schubnell}, {Sevilla}, {Smith}, {Stalder}, {Stark}, {Strazzullo}, {Suchyta}, {Swanson},
  {Tarle}, {Thaler}, {Thomas}, {Tucker}, {Vikram}, {von der Linden}, {Walker}, {Wechsler}, {Wester}, {Zenteno}, \& {Ziegler}}]{Saro15}
{Saro}, A., {Bocquet}, S., {Rozo}, E., {et~al.} 2015, \mnras, 454, 2305, \dodoi{10.1093/mnras/stv2141}

\bibitem[{{Sereno} \& {Zitrin}(2012)}]{Sereno12}
{Sereno}, M., \& {Zitrin}, A. 2012, \mnras, 419, 3280, \dodoi{10.1111/j.1365-2966.2011.19968.x}

\bibitem[{{Serra} {et~al.}(2011){Serra}, {Diaferio}, {Murante}, \& {Borgani}}]{Serra11}
{Serra}, A.~L., {Diaferio}, A., {Murante}, G., \& {Borgani}, S. 2011, \mnras, 412, 800, \dodoi{10.1111/j.1365-2966.2010.17946.x}

\bibitem[{{Sif{\'o}n} {et~al.}(2016){Sif{\'o}n}, {Battaglia}, {Hasselfield}, {Menanteau}, {Barrientos}, {Bond}, {Crichton}, {Devlin}, {D{\"u}nner}, {Hilton}, {Hincks}, {Hlozek}, {Huffenberger}, {Hughes}, {Infante}, {Kosowsky}, {Marsden}, {Marriage}, {Moodley}, {Niemack}, {Page}, {Spergel}, {Staggs}, {Trac}, \& {Wollack}}]{Sifon16}
{Sif{\'o}n}, C., {Battaglia}, N., {Hasselfield}, M., {et~al.} 2016, \mnras, 461, 248, \dodoi{10.1093/mnras/stw1284}

\bibitem[{{Simet} {et~al.}(2017){Simet}, {McClintock}, {Mandelbaum}, {Rozo}, {Rykoff}, {Sheldon}, \& {Wechsler}}]{Simet17}
{Simet}, M., {McClintock}, T., {Mandelbaum}, R., {et~al.} 2017, \mnras, 466, 3103, \dodoi{10.1093/mnras/stw3250}

\bibitem[{{Stalder} {et~al.}(2013){Stalder}, {Ruel}, {{\v{S}}uhada}, {Brodwin}, {Aird}, {Andersson}, {Armstrong}, {Ashby}, {Bautz}, {Bayliss}, {Bazin}, {Benson}, {Bleem}, {Carlstrom}, {Chang}, {Cho}, {Clocchiatti}, {Crawford}, {Crites}, {de Haan}, {Desai}, {Dobbs}, {Dudley}, {Foley}, {Forman}, {George}, {Gettings}, {Gladders}, {Gonzalez}, {Halverson}, {Harrington}, {High}, {Holder}, {Holzapfel}, {Hoover}, {Hrubes}, {Jones}, {Joy}, {Keisler}, {Knox}, {Lee}, {Leitch}, {Liu}, {Lueker}, {Luong-Van}, {Mantz}, {Marrone}, {McDonald}, {McMahon}, {Mehl}, {Meyer}, {Mocanu}, {Mohr}, {Montroy}, {Murray}, {Natoli}, {Nurgaliev}, {Padin}, {Plagge}, {Pryke}, {Reichardt}, {Rest}, {Ruhl}, {Saliwanchik}, {Saro}, {Sayre}, {Schaffer}, {Shaw}, {Shirokoff}, {Song}, {Spieler}, {Stanford}, {Staniszewski}, {Stark}, {Story}, {Stubbs}, {van Engelen}, {Vanderlinde}, {Vieira}, {Vikhlinin}, {Williamson}, {Zahn}, \& {Zenteno}}]{Stalder13}
{Stalder}, B., {Ruel}, J., {{\v{S}}uhada}, R., {et~al.} 2013, \apj, 763, 93, \dodoi{10.1088/0004-637X/763/2/93}

\bibitem[{{The} \& {White}(1986)}]{The86}
{The}, L.~S., \& {White}, S.~D.~M. 1986, \aj, 92, 1248, \dodoi{10.1086/114258}

\bibitem[{{Vikhlinin} {et~al.}(2009){Vikhlinin}, {Burenin}, {Ebeling}, {Forman}, {Hornstrup}, {Jones}, {Kravtsov}, {Murray}, {Nagai}, {Quintana}, \& {Voevodkin}}]{Vikhlinin09}
{Vikhlinin}, A., {Burenin}, R.~A., {Ebeling}, H., {et~al.} 2009, \apj, 692, 1033, \dodoi{10.1088/0004-637X/692/2/1033}

\bibitem[{{Voit}(2005)}]{Voit05}
{Voit}, G.~M. 2005, Reviews of Modern Physics, 77, 207, \dodoi{10.1103/RevModPhys.77.207}

\bibitem[{{Wilson} {et~al.}(2009){Wilson}, {Muzzin}, {Yee}, {Lacy}, {Surace}, {Gilbank}, {Blindert}, {Hoekstra}, {Majumdar}, {Demarco}, {Gardner}, {Gladders}, \& {Lonsdale}}]{Wilson09}
{Wilson}, G., {Muzzin}, A., {Yee}, H.~K.~C., {et~al.} 2009, \apj, 698, 1943, \dodoi{10.1088/0004-637X/698/2/1943}

\bibitem[{{Yahil} \& {Vidal}(1977)}]{Yahil77}
{Yahil}, A., \& {Vidal}, N.~V. 1977, \apj, 214, 347, \dodoi{10.1086/155257}

\bibitem[{{Yang} {et~al.}(2005){Yang}, {Mo}, {van den Bosch}, \& {Jing}}]{Yang05}
{Yang}, X., {Mo}, H.~J., {van den Bosch}, F.~C., \& {Jing}, Y.~P. 2005, \mnras, 356, 1293, \dodoi{10.1111/j.1365-2966.2005.08560.x}

\bibitem[{{Yang} {et~al.}(2007){Yang}, {Mo}, {van den Bosch}, {Pasquali}, {Li}, \& {Barden}}]{Yang07}
{Yang}, X., {Mo}, H.~J., {van den Bosch}, F.~C., {et~al.} 2007, \apj, 671, 153, \dodoi{10.1086/522027}

\bibitem[{{Zabludoff} {et~al.}(1990){Zabludoff}, {Huchra}, \& {Geller}}]{Zabludoff90}
{Zabludoff}, A.~I., {Huchra}, J.~P., \& {Geller}, M.~J. 1990, \apjs, 74, 1, \dodoi{10.1086/191492}

\bibitem[{{Zenteno} {et~al.}(2016){Zenteno}, {Mohr}, {Desai}, {Stalder}, {Saro}, {Dietrich}, {Bayliss}, {Bocquet}, {Chiu}, {Gonzalez}, {Gangkofner}, {Gupta}, {Hlavacek-Larrondo}, {McDonald}, {Reichardt}, \& {Rest}}]{Zenteno16}
{Zenteno}, A., {Mohr}, J.~J., {Desai}, S., {et~al.} 2016, \mnras, 462, 830, \dodoi{10.1093/mnras/stw1649}

\end{thebibliography}
\bibliographystyle{aasjournal}

\end{document}